\documentclass[fleqn,usenatbib]{mnras}

\usepackage{newtxtext,newtxmath}

\usepackage[T1]{fontenc}
\usepackage{ae,aecompl}

\usepackage{graphicx}	
\usepackage{amsmath}	
\usepackage{amssymb}	

\newcommand{\update}[1]{{\color[rgb]{0.0, 0.0, 0.0}#1}}

\title[CALIFA MDFs]{The stellar metallicity distribution function of galaxies in the CALIFA survey}

\author[A. Mej\'ia-Narv\'aez et al.]{%
A. Mej\'ia-Narv\'aez$^{1}$\thanks{E-mail: amejia@astro.unam.mx (AMN)},
S.~F. S\'anchez,$^{1}$
E.~A. D. Lacerda,$^{1}$
L. Carigi,$^{1}$
L. Galbany,$^{2}$\newauthor
B. Husemann,$^3$
and R. Garc\'ia-Benito$^{4}$
\\
$^{1}$Instituto de Astronom\'ia, Universidad Nacional Aut\'onoma de M\'exico, A. P. 70-264, C.P. 04510 M\'exico, D.F., M\'exico\\
$^{2}$Departamento de F\'isica Te\'orica y del Cosmos, Universidad de Granada, E-18071 Granada, Spain\\
$^{3}$Max-Planck-Institut f\"ur Astronomie, K\"onigstuhl 17, D-69117 Heidelberg, Germany\\
$^{4}$Instituto de Astrof\'isica de Andaluc\'ia (IAA/CSIC), Glorieta de la Astronom\'ia s/n Aptdo. 3004, E-18080 Granada, Spain
}

\date{Accepted XXX. Received YYY; in original form ZZZ}

\pubyear{2019}

\begin{document}
\label{firstpage}
\pagerange{\pageref{firstpage}--\pageref{lastpage}}
\maketitle

\begin{abstract}
We present a novel method to retrieve the chemical structure of galaxies using integral field spectroscopy data through the stellar Metallicity Distribution Function (MDF). \update{This is the probability distribution of observing stellar populations having a metallicity $Z$.} We apply this method to a set of $550$ galaxies from the CALIFA survey. We present the behaviour of the MDF as a function of the morphology, the stellar mass and the radial distance. We use the stellar metallicity radial profiles retrieved as the first moment of the MDF, as an internal test for our method. The gradients in these radial profiles are consistent with the known trends: they are negative in massive early-type galaxies and tend to positive values in less massive late-type ones. We find that these radial profiles may not convey the complex chemical structure of some galaxy types. Overall, low mass galaxies ($\log{M_\star/\text{M}_{\sun}}\leq10$) have broad MDFs ($\sigma_Z\sim1.0\,$dex), with unclear dependence on their morphology. However this result is likely affected by under-represented bins in our sample. On the other hand, massive galaxies ($\log{M_\star/\text{M}_{\sun}}\geq11$) have systematically narrower MDFs ($\sigma_Z\leq0.2\,$dex). We find a clear trend whereby the MDFs at $r_k/R_e>1.5$ have large variance. This result is consistent with sparse SFHs in medium/low stellar density regions. We further find there are multi-modal MDFs in the outskirts ($\sim18\,$per cent) and the central regions ($\sim40\,$per cent) of galaxies. This behaviour is linked to a fast chemical enrichment during early stages of the SFH, along with the posterior formation of a metal-poor stellar population.
\end{abstract}




\section{Introduction}

Chemical enrichment of the interstellar medium (ISM), depends mainly on the chemical elements synthesized during the stellar evolution. \update{As gas cools down at the center of dark matter halos, star formation takes place \citep[e.~g.,][]{Agertz2009}}. Eventually, the newly formed stars evolve and pollute the ISM with their chemical by-products, predominantly during their last evolutionary phases (stellar winds in the AGB phase and/or SNe events). The production of new chemical elements and the lifetime of any star, depends strongly on its initial mass and metallicity, $Z$ \citep{Yates2013}. Consequently, the timescales of the ISM enrichment by different elements are functions of the initial stellar mass distribution during star formation events \citep[e.~g.,][]{Poulhazan2018, Jerabkova2018}. A physically relevant trait of the following generations of stars born from this polluted ISM, is that they are chemically enriched with respect to their progenitors.

However the chemical evolution does not occur in a linear manner. It is well known from the stellar evolution theory that more massive stars ($>10\,\text{M}_{\sun}$) evolve more quickly \citep[$\sim10\,$Myr, e.~g.,][]{Portinari1998}. Therefore, these stars are the first to enrich the ISM with heavy chemical species. On the other hand, the intermediate and low mass stars ($<10\,\text{M}_{\sun}$), lock the chemical products of past generations for longer times \citep[$\sim 1\,$---$\,10\,$Gyr, e.~g.,][]{Marigo2001, Thielemann2003}. An immediate consequence of such differential and time-delayed enrichment, is that second generation stars carry the chemical imprint left by the dying massive stars, whilst there are always living low mass stars with the chemical composition of the ISM across the star formation history (SFH).

From the observational point of view, we know the stellar populations of different ages and initial chemical composition, encode the chemical evolution fossil record in several regions of the integrated spectral energy distribution of galaxies. Therefore, we can virtually recover such Chemical Evolution Histories (ChEHs) of the ISM through spectral fitting methods \citep{Panter2008, ValeAsari2009}. Indeed, studies of the chemical evolution of unresolved galaxies have unravelled the \emph{global} interplay between the stellar mass, the gas mass fraction and the stellar metallicity. Essentially low-mass galaxies ($\log{M_\star/\text{M}_{\sun}}<10$) exhibit larger fractions of gas, a larger scatter in stellar metallicity and a mild bias towards low metallicity, than more massive counterparts. On the other hand, high mass galaxies ($\log{M_\star/\text{M}_{\sun}}>10$) concentrate in a locus of lower gas fraction, higher metallicity and little scatter (e.~g., \cite{Tremonti2004, Brinchmann2004, Gallazzi2008} and see \cite{Blanton2009}, for a review).

More recently, spatially-resolved surveys of galaxies have revealed the local character of such relations \citep{Rosales-Ortega2012, Sanchez2013, Wuyts2013, Cano-Diaz2016}. There are two main drivers defining the timescales of the resolved stellar population properties: \update{\textit{i)} the total stellar mass which essentially traces the gravitational potential well of the galactic dark matter halo \citep[e.~g.,][]{Matteucci1994, Thomas2005}}, and \textit{ii)} the stellar mass surface density which traces the local gravitational potential \citep[e.~g.,][]{GonzalezDelgado2014a}. In this sense, it appears that in more massive galaxies large scale phenomena ($\sim0.1\,$---$\,10\,$Mpc, dark halo relaxation, gas cooling, etc.) takes place in shorter timescales than in less massive ones. However, at smaller spatial scales ($\sim1\,\text{kpc}$), even though the timescales of physical phenomena like star formation, chemical evolution are still bound by the global mass of the galaxies, the dominant factor is the local mass density: denser regions evolve more quickly than less dense ones \citep[e.~g.,][]{Perez2013, Garcia-Benito2017}.

Although most of our knowledge on the stellar chemical structure of galaxies comes from the study of its radial profiles \citep[e.~g.,][]{GonzalezDelgado2014, Goddard2017}, it remains to be seen if those are indeed representative of the underlying metallicity distribution across the galaxies extent. As a matter of fact, a plethora of studies jointly demonstrate the complexities of the resolved chemical distributions in the Milky Way neighborhood \citep[e.~g.,][]{Kirby2011, Hayden2015, Martinez-Medina2017, Frankel2019, Escala2020}. In this paper we introduce a novel method to study the fully \emph{resolved} stellar chemical structure of galaxies. We investigate the results of our method on a sample of the extended CALIFA (Calar Alto Legacy Integral Field Area) survey \citep[][S\'anchez et al., in prep.]{Sanchez2012}. \update{As starting point, we recover the well-known stellar metallicity radial profiles out of the stellar Metallicity Distribution Functions (MDFs). Essentially, the MDF describes the probability distribution of observing stellar populations having metallicity $Z$. Then, we compare the physical content between those radial profiles and the integrated version of these distributions.} Finally we analyze the spatially resolved MDF for all the galaxies in our sample to uncover patterns as a function of morphological class and total stellar mass, following previous works in the same line \citep[e.~g.,][]{GonzalezDelgado2015}.

This paper is organized as follows. In section \S~\ref{sec:sample} we describe the sample, while the method used to extract the physical properties from the the integral field spectroscopy data sets is briefly described in \S~\ref{sec:spectral-fitting}. The method for reconstructing the MDFs is introduced in \S~\ref{sec:mdf-method}. In section \S~\ref{sec:metallicity-profiles} we show the radial profiles of the stellar metallicity of galaxies in our sample and discuss them, using previous results as a benchmark. In section \S~\ref{sec:mdf-global} we present the MDF of CALIFA galaxies in two different ways: first we present the integrated MDFs across the galaxies' extent as a function of morphological class and stellar mass. We compare its physical value to the radial profiles presented in the previous section. Then, in \S~\ref{sec:mdf-resolved}, we introduce the spatially resolved MDFs in the same morphological class and stellar mass bins and elaborate on the observed trends. Finally we discuss our findings in section \S~\ref{sec:discussion} and conclude in section \S~\ref{sec:conclusions}. \update{Throughout this study we assume the standard $\Lambda$ Cold Dark Matter cosmology with the parameters: H$_0$=71 km/s/Mpc, $\Omega_M$=0.27, $\Omega_\Lambda$=0.73.}

\section{Sample}\label{sec:sample}

In this paper we adopt the extended CALIFA sample, comprising: \textit{(i)} all galaxies observed with the V500 setup extracted from the original survey \citep[][]{Sanchez2012}, and \textit{(ii)} additional observations of under-represented galaxy types observed with the same setup in different extended surveys, in particular the PISCO survey \citep[PMAS/Ppak Integral-field Supernova hosts COmpilation,][]{Galbany2018}. The final sample comprises $\sim900$ galaxies spanning all morphological classes. The spectroscopic data for these galaxies was obtained using the Potsdam Multi-Aperture Spectrophotometer \citep[PMAS,][]{Roth2005} at the $3.5\,\text{m}$ telescope of the Calar Alto Observatory (CAHA). The spectroscopic configuration (V500) allows gathering galaxy spectra that spans the optical range from $\sim3700\,$\AA~to $7000\,$\AA~ at a mean resolution power of $R\sim850$, suitable for retrieving the stellar metallicity out of spectral fitting methods. The spatially-resolved characteristic of the observations was achieved in the PPak mode \citep{Verheijen2004}. This setting comprises $382$ fibers of $2.7''$ aperture (diameter) covering a total $74\times64''$ field-of-view, $331$ of which are arranged in a compact fiber-bundle \citep[FoV,][]{Kelz2005}. A three dithering scheme was performed to achieve a full sampling of the complete FoV. After reducing the data, the final dataset consists of individual cubes with two dimensions sampling the spatial coordinates within the FoV and the third one sampling the observed spectral range. Reduction was performed using version 2.2 of the Data Reduction Pipeline \citep{Husemann2013, Garcia-Benito2015, Sanchez2016b}. The final datacubes have an average spatial resolution of FWHM$\sim2.5\arcsec$. At the typical redshift of the sample, the physical resolution is $0.8\,$kpc, so that at each spaxel ($\sim300\,$pc) the survey is actually sampling complex stellar mixtures. For more details on the sample, the reader is referred to \cite{Walcher2014}.

\begin{figure}
\includegraphics[scale=0.45]{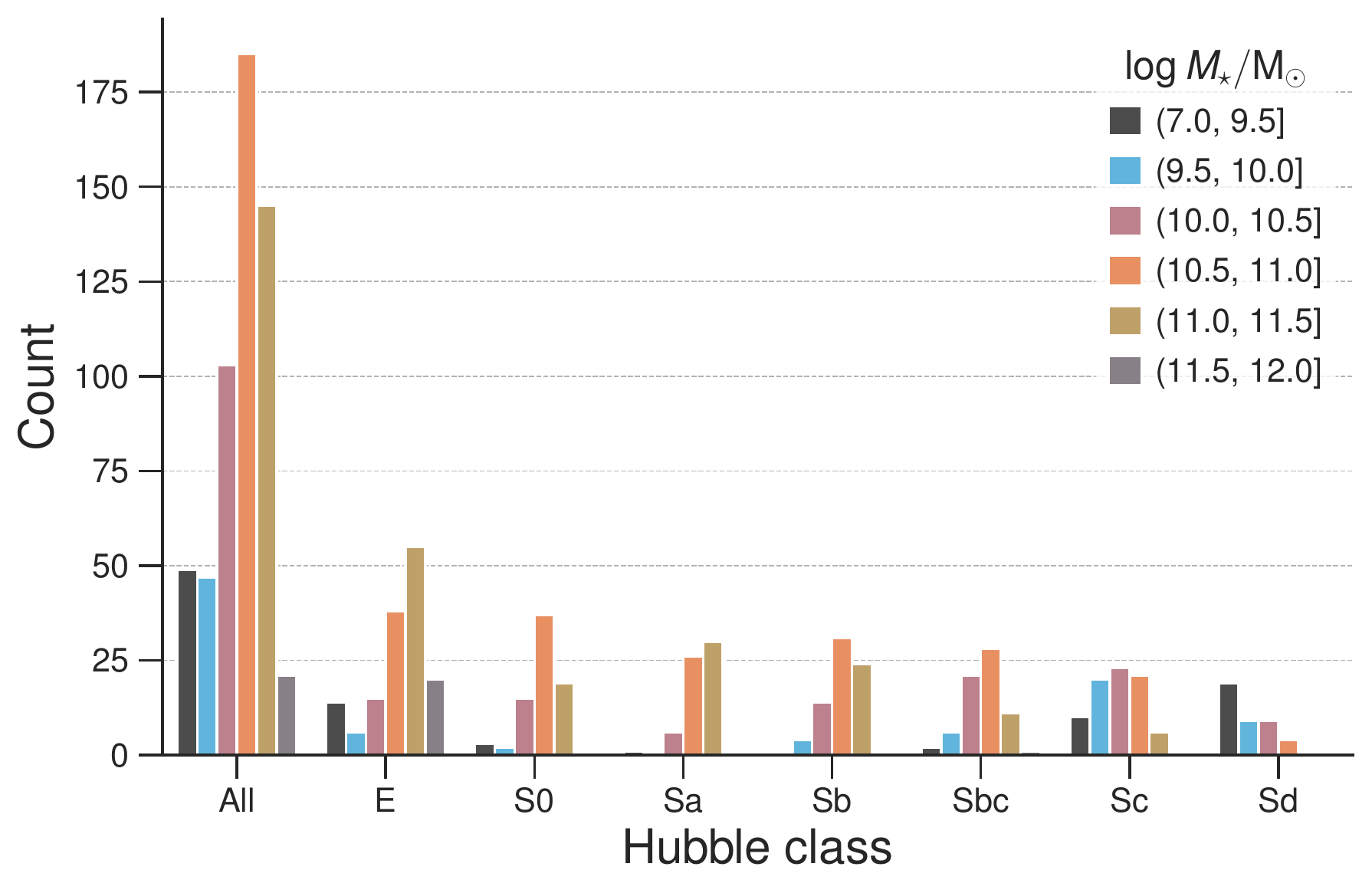}
\caption{\update{The main sample of galaxies adopted in this paper is shown in histograms segregated by morphological class (derived by visual inspection) and stellar mass bins (computed using the \textsc{FIT3D})}. Clearly some morphologies have under-sampled stellar mass bins (color coded) and only E-type galaxies have all mass bins represented. Overall the trends show that later galaxy types transit from higher stellar masses towards lower stellar masses.}
\label{fig:sample-distributions}
\end{figure}
From this original sample we selected those objects more suitable for the current analysis by applying the following criteria: \textit{(i)} only data cubes visually inspected and flagged as clean make it to the sample ($859$ galaxies),\footnote{FoV covered by galaxy with a small fraction of bad pixels and no important contaminants such as bright stars.} \textit{(ii)} in order to avoid inclination effects on our estimates, we filter out disc-shaped galaxies with inclinations above $70\,\deg$, so that only those galaxies closer to being face-on are part of our analysis ($633$ galaxies), \textit{(iii)} we keep galaxies that are isolated and show no sign of being merger remnants after visual inspection ($594$ galaxies), \textit{(iv)} we remove galaxies that host an AGN following the recipe by \cite{Lacerda2020}, since we are only interested in galaxies dominated by their stellar component ($573$ galaxies), and \textit{(v)} we remove galaxies classified with morphologies Sm and I since their galactocentric distance and inclinations are unclear. \update{The final sample comprises $550$ galaxies, spanning a wide range of morphological classes and stellar masses. In Fig.~\ref{fig:sample-distributions}, we show the distribution of galaxies in such sample, segregated by morphological class and stellar mass. The morphological classification was performed by the Stellar Populations and Chemical Evolution groups at Instituto de Astronom\'ia~--~UNAM, through visual inspection. The stellar masses are estimated using the spectral analysis pipeline described in the next section.}

\section{Methods}

Before any science can be done out of the reduced spatially-resolved spectroscopy data, a series of pre-processing, physical properties extraction and post-processing/compression steps need to be taken. In this paper we have adopted the \textsc{Pipe3D} pipeline \citep{Sanchez2016}, in the core of which lies the spectral fitting algorithm \textsc{FIT3D} \citep{Sanchez2016a}. In the next section we elaborate briefly on the aspects of such procedures that are most relevant to this work.

\subsection{\textsc{FIT3D}: spectral modelling algorithm}\label{sec:spectral-fitting}

\begin{figure*}
\includegraphics[scale=0.5]{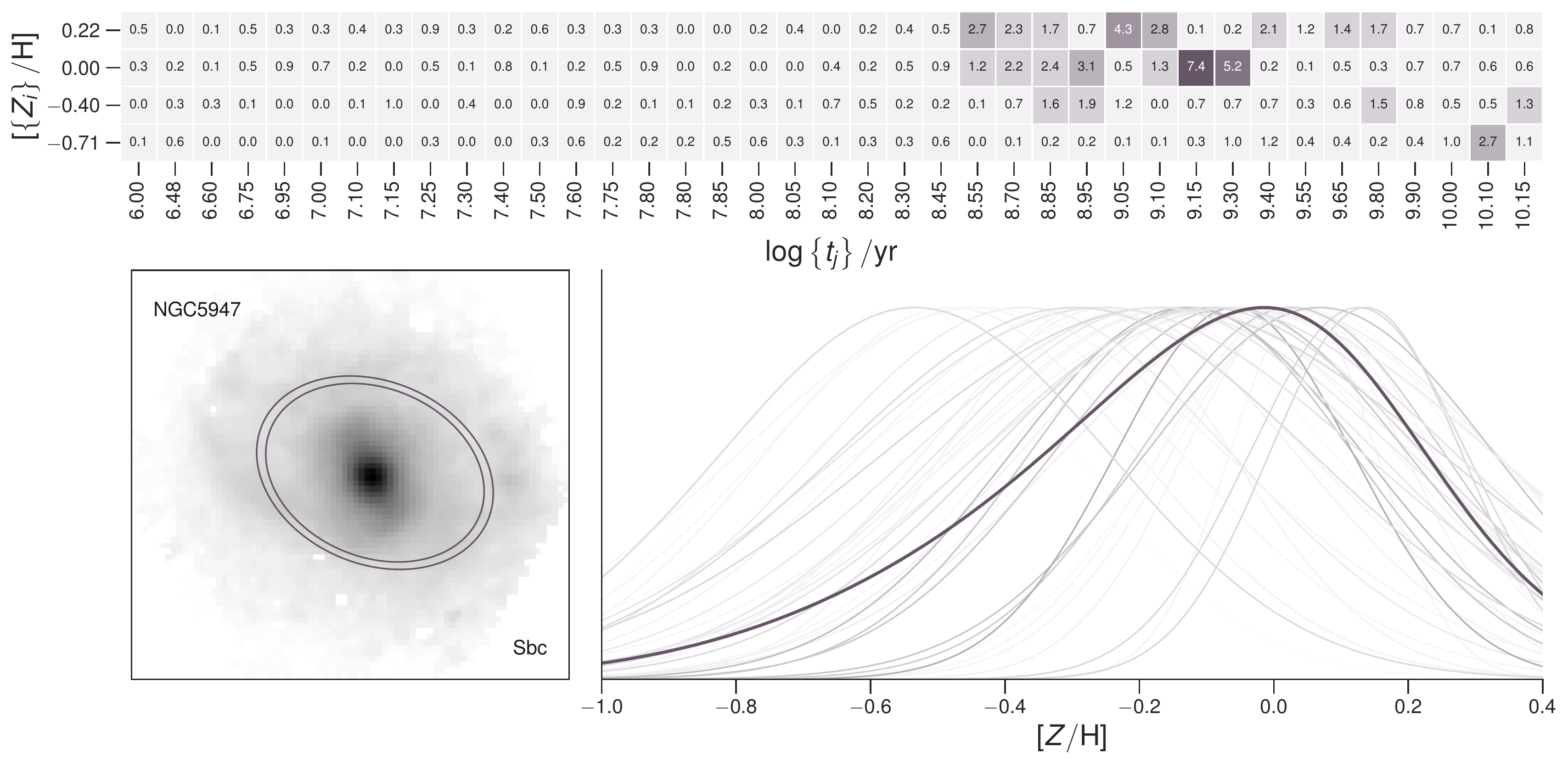}
\caption{Scheme of the MDF reconstruction for NGC 5947 within an arbitrary radial bin $r_k\sim8\,$kpc. \textit{(i)} Top panel, the percentage contribution in light of each SSP model in the parameter space defined by $\{Z_i\}$ and $\{t_j\}$; \textit{(ii)} bottom-left panel, the optical image of the galaxy along with the ellipse enclosing the region where the SFH and ChEH was computed for this example; bottom-right panel, the kernels used in grey-scale according to their contribution in light to the resulting MDF (thick line). We highlight that the kernels may have different spreads, as opposed to the conventional definition of the KDE.}
\label{fig:mdf-demonstration}
\end{figure*}

The \textsc{FIT3D} code implements a multi-component non-parametric spectral fitting algorithm specially suited for the analysis of CALIFA datacubes \citep{Sanchez2006}, although it has been extensively used in other surveys 
(e.~g., PINGS, \cite{Rosales-Ortega2010}; MaNGA, \cite{Ellison2018}; SAMI, \cite{Sanchez2019a}; AMUSSING++, \cite{Lopez-Coba2020N}). The non-parametric character comes from the non-assumption of the functional shape of the SFH and the ChEH, while the multi-component character arises from the fact that both, the stellar and the interstellar medium are disentangled from the observed spectrum. The procedure for running such decomposition can be summarized as follows:
\begin{description}
\item[\textit{i) The non-linear analysis:}] in the initial part of the fitting procedure, the parameter space comprising the line-of-sight velocity dispersion ($\sigma_\star$), the systemic velocity ($v_\star$) and the interstellar dust extinction ($A_V$), is explored using a linear combination of a minimum set of simple stellar populations (SSP) spectra with the objective of minimizing the merit function $\chi^2(\sigma_\star,v_\star,A_V)$. The properties of such set ingredients were chosen so that the degeneracies between resulting physical properties is kept at minimum.
\item[\textit{ii) Ionized gas/star emission decomposition:}] the ionized gas emission modelling assumes Gaussian profiles, taking into account the relationships (e.~g., kinematics, line flux ratios) among the several emission lines in the optical spectrum. Therefore, the parameter space is sampled in bulks of lines that are physically tied to one another, producing faster and more plausible results. Once the gas emission model is fit, its spectral contribution is subtracted from the observed spectrum, so that $f_\lambda^\star = f_\lambda^\text{obs} - f_\lambda^\text{gas}$ is the pure stellar emission. Finally, this stellar component, is further decomposed into a linear combination of simple stellar populations by mapping a grid of ages and stellar metallicities into the $f_\lambda^\star$ spectrum, as follows:
\begin{equation}\label{eq:stellar-model}
f_\lambda^\text{model} = \sum_{ij}\,w_{ij}\,f_\lambda^\text{SSP}(Z_i; t_j),
\end{equation}
where the non-linear parameters derived from the previous step are already included in $f_\lambda^\text{SSP}$, the monochromatic luminosity predicted by the SSP of age $t_j$ and metallicity $Z_i$. The weights $w_{ij}$ represent the light contribution of each SSP to the target spectrum.
\end{description}

In this study we use the GSD156 library of SSP models from \cite{CidFernandes2013}, spanning $39$ ages in the range $\{t_j\}=1\,$Myr~---~$14.1\,$Gyr, and the $4$ stellar metallicities $\{Z_i\}=0.004$, $0.008$, $0.019$, $0.032$. We assume the initial mass function from \cite{Salpeter1955}.
This library of model ingredients has been implemented in a number of studies \cite[e.~g.,][]{Perez2013, Lin2019, Lacerda2020}. Throughout this study we adopt the metallicity scale defined by $[Z/\text{H}] \equiv \log{Z/Z_{\sun}}$, where we assume $Z_{\sun}=0.019$. The spatially-resolved properties are presented in radial bins of width $\delta r_k/R_e=0.6$, centred at $\{r_k/R_e\}=\{0.3, 0.9, 1.5, 2.1, 2.7, 3.3\}$, where we scale the radial dimension of each galaxy in our sample using its effective radius, $R_e$.

Overall the \textsc{FIT3D} code implements the minimization of the merit function $\chi^2$ across the parameter space, yielding point estimates of all physical properties. However, to provide an internal estimation of the errors in these estimates, each step of the fit is repeated in a Monte Carlo fashion, randomly perturbing the original target spectrum. The best estimate is then reported as the average and the associated uncertainty as the standard deviation. Both are computed over the value derived from each random perturbation fitting. Finally, the post-processing consists in building the maps of each physical property from the \textsc{FIT3D} spectral decomposition, across the galaxy extent. In this paper we are mainly concerned on the SFH and the ChEH, these are, essentially, the maps of the weights $w_{ij}$. In the next section we develop on the statistical construction needed to retrieve the MDF out of these particular data products.

\subsection{Statistical reconstruction of the MDF}\label{sec:mdf-method}

In general the implementation of the spectral fitting method of spatially-resolved data sets would yield the average of some stellar physical property at all polar coordinate $(r,\theta)$ across the target galaxy extent. However, let us assume azimuthal symmetry to recover the radial structures as in previous works \citep[e.~g.][]{Sanchez-Blazquez2014, GonzalezDelgado2016}. If this assumption does not hold for the chemical structure of the galaxies, we can foresee that the recovered MDF will manifest non-Gaussian behaviours such as asymmetries, multi-modes, etc.. It is worth to point out this would not be the case for the radial profiles, since the average would conceal the underlying complexities present in the chemical structure. From first principles, we could describe the \emph{resolved} chemical structure of the galaxies as a function of the radial coordinate $r$, as follows:
\begin{equation}\label{eq:mdf-theoretical}
\update{\text{MDF}(r) = \int P(Z, t\,|\,r)\,\text{d}t,}
\end{equation}
where $P$ is the probability of the target galaxy having a stellar population characterized by mixtures of stellar metallicities $Z$ and ages $t$ at radius $r$. However the spectral fitting method we adopted samples only the maximum likelihood in the parameter space, so that we only have $\hat{\mathcal{L}}(\{f_\lambda^\text{obs}\}\,|\,\{Z_i\}, \{t_j\}, r_k)$. This is essentially the sampling of the most likely SFH and ChEH at each radial bin, $r_k$. Furthermore, the metallicity grid is coarsely sampled. Therefore, the \emph{a posteriori} distribution is, by construction, under-sampled. For these reasons, in order to reliably recover the $\text{MDF}(r_k)$ we will implement a density estimation method.

The Kernel Density Estimator \citep[KDE;][]{Rosenblatt1956, Parzen1962} is a non-parametric way to uncover the underlying probability density function (PDF) for a given sample. Contrary to the conventional approach of inferring a functional form given a histogram, the KDE poses some advantages. First, there is no need to specify the bin positions. Each sampled point is considered a bin. Second, the width of each bin can be directly inferred from the uncertainties in each sampled point. Finally, if the kernel $K$ takes the form of a known PDF, the KDE can be written simply as the weighted average, as follows:
\begin{equation}\label{eq:kde-definition}
\text{KDE}=\frac{1}{Nh}\sum_{j=1}^{N}\,K(\delta_j/h),
\end{equation}
where $\delta_j = \mu_{Z,j} - Z$ is the distance between some arbitrary point in the PDF support, $Z$, and a point in the sample, $\mu_{Z,j}$, and $j=1,\ldots,N$. $h$ is known as the bandwidth and is a free parameter in the KDE. Typically, the choice of $K$ is such that it decreases with $|\delta_j|$. The choice of the bandwidth value is such that the KDE is a compromise between the relevant features (for instance bimodalities) and little noise contributions to the KDE.

To adjust the KDE method for the purpose of this study, we need to make some additional considerations on the sample to fully encompass its physical meaning. By definition of the spectral fitting method adopted, the SFH and the ChEH (described by the sets $\{t_j\}$ and $\{Z_i\}$, and the weights $w_{ij}$) are related to each other so that for each stellar age $t_j$ there is a set $\{Z_i\}$. Therefore, the (azimuthal-averaged) stellar metallicity at some point in time $t_j$ and at any radial location $r_k$ within each galaxy can be expressed as:
\begin{equation}\label{eq:kernel-mu}
\mu_{Z,j} \equiv \mu_Z(t_j) = \frac{\sum_{i=1}^{N_Z}\,w_{ij}\,Z_i}{\sum_{i=1}^{N_Z}\,w_{ij}},
\end{equation}
where $N_Z$ is the number of stellar metallicities in the model grid. The typical (weighted) standard deviation from this average is:
\begin{equation}\label{eq:kernel-sigma}
\sigma_{Z,j} \equiv \sigma_Z(t_j) = \sqrt{\frac{\sum_{i=1}^{N_Z}\,w_{ij}\,\left(Z_i - \mu_Z\right)^2}{(M_Z-1)/M_Z\,\sum_{i=1}^{N_Z}\,w_{ij}}},
\end{equation}
where $M_Z$ is the number of non-zero weights $w_{ij}$ along $i$.\footnote{It is worth to note that the standard deviation in Eq.~\eqref{eq:kernel-sigma} is not meant to be an estimation of the error in the stellar metallicity recovered, but a measure of the mixture of $Z$ at a given point in time $t_j$ and radial position $r_k$.} This way, our KDE definition in Eq.~\eqref{eq:kde-definition} can be turned into the MDF at each radial position, $r_k$, as follows:
\begin{equation}\label{eq:mdf-definition}
\text{MDF}(r_k) = \frac{1}{N_th}\frac{\sum_{j=1}^{N_t}\,w_j\,K(\delta_j/h)}{\sum_{j=1}^{N_t}\,w_j},
\end{equation}
where $N_t$ is the number of stellar ages and we are assuming a Gaussian kernel expressed as:
\begin{equation}\label{eq:mdf-kernel}
K(\delta_j/h) \equiv \frac{1}{\sqrt{2\uppi}\sigma_{Z,j}}\exp{\left[-\frac{\left(\mu_{Z,j} - Z\right)^2}{2h\sigma_{Z,j}^2}\right]}.
\end{equation}

The Gaussian shape of the kernel is an \emph{ad hoc} assumption, however it still reflects the fact that uncertainties in stellar metallicity estimates from spectral fitting often show such behaviour \citep{CidFernandes2014,Sanchez2016a}. Furthermore, it provides a smooth and known parametrization for the MDF which is consistent with our current understanding of the MDF shape for the Milky Way neighborhood \citep[e.~g.,][]{Hayden2015, Escala2020, Kirby2020}. The remaining free parameter, the bandwidth $h$, is set to unity for the rest of this study. As mentioned before, this parameter is set to retrieve smooth, yet informative KDE. However, the intrinsic spread of the kernel defined in Eq.~\eqref{eq:kernel-sigma}, which will depend on $\{t_j\}$, provides a physical insight on the underlying stellar metallicity distribution for a given stellar age, at a given radius. By setting $h=1$, we allow for such insight be reflected in the MDF.

In Fig.~\ref{fig:mdf-demonstration} we highlight the main stages to build the MDF defined in Eq.~\eqref{eq:mdf-definition}. The top panel shows the distribution of weights $w_{ij}$ (in percentages) for NGC $5947$ within the annulus around $\sim8\,$kpc, as shown in the bottom left panel. In the bottom right panel, the kernels as defined in Eq.~\eqref{eq:mdf-kernel} are drawn grey-shaded according to their contribution to the final MDF, also drawn with a thick dark line. We implemented this tailored solution, i.~e., the KDE-like, since in our opinion it provides the best physical interpretations on the chemical structure of galaxies, with the current data. \update{In Appendix~\ref{app:reliability} we test for the reliability of the method presented in \S~\ref{sec:mdf-method} using toy simulations of SFH and ChEHs.}

\section{Results}\label{sec:results}

The MDF estimated by Eq.~\eqref{eq:mdf-definition} can be physically interpreted as the probability of \emph{observing} a stellar population having a metallicity $Z$ at some radial bin $r_k$. This is the conditional probability on $r_k$. Under this circumstances, the rules of probabilities state that the integrated MDF across some galaxy extent should be expressed as the summation:
\begin{equation}\label{eq:mdf-galaxy}
\text{MDF}_\text{galaxy} \propto \sum_{k=1}^{N_r}\,\text{MDF}(r_k).
\end{equation}
This is equivalent to marginalize the MDF, as defined in Eq.~\eqref{eq:mdf-theoretical}, along the radial dimension. On the other hand, the MDFs within a given combination of morphological and/or stellar mass bin are in fact independent from one another (assuming isolation), so that the MDF can be written, using the product rule of probabilities, as:
\begin{equation}\label{eq:mdf-bin}
\text{MDF}_\text{bin}(r_k) \propto \prod_{m=1}^{N_\text{bin}}\,\text{MDF}_m(r_k),
\end{equation}
where $m=1,\ldots,N_\text{bin}$ runs over morphological and/or stellar mass bins. In Appendix~\ref{app:probabilities} we expand on the algebra of probabilities starting with the definition in Eq.~\eqref{eq:mdf-theoretical} in order to justify Eqs.~\eqref{eq:mdf-galaxy} and \eqref{eq:mdf-bin}.

Before we introduce the MDF retrieved using the method explained above, in the next section we revisit the radial stellar metallicity profiles of the galaxies in the sample defined in \S~\ref{sec:sample}. If our method for describing the underlying MDF is reliable, we should be able to retrieve the same trends in the literature, within the known uncertainties.

\subsection{Recovering the stellar metallicity profiles}\label{sec:metallicity-profiles}

\begin{figure*}
\includegraphics[scale=0.45]{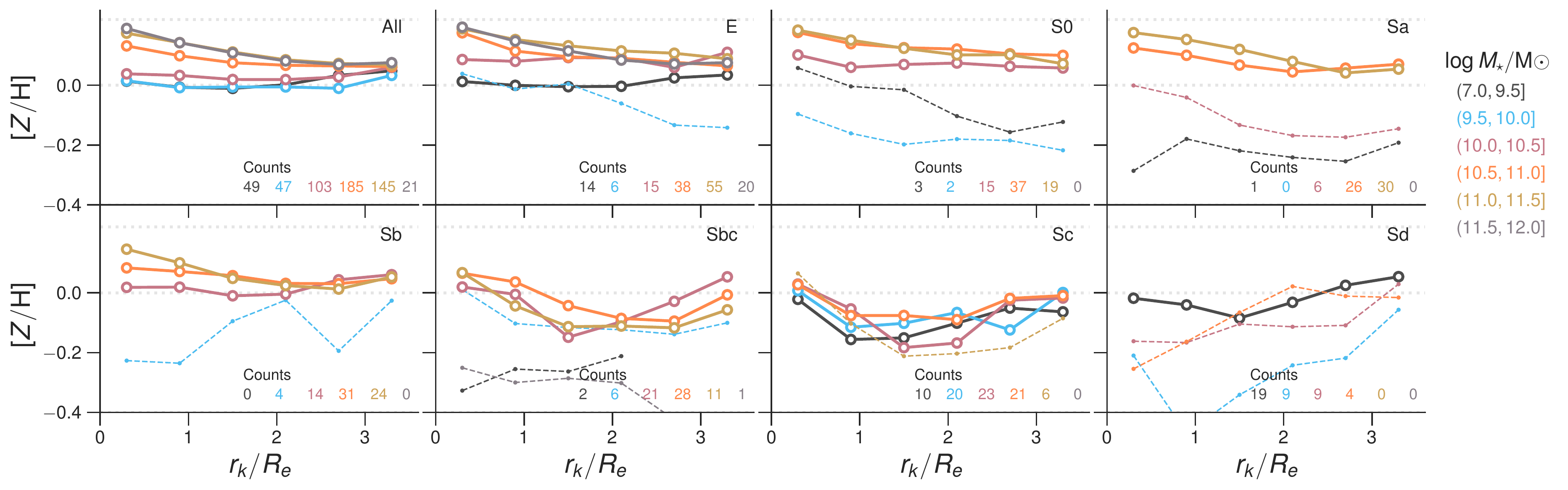}
\caption{The stellar metallicity radial profiles obtained from the first moment of the resolved MDFs are shown for the sample of galaxies defined in \S~\ref{sec:sample}, segregated in the same fashion as in Fig.~\ref{fig:sample-distributions}, are shown as points located at the radial bins $\{r_k\}$. The joining lines have the sole intent to guide the eye and hold no physical meaning. The number of galaxies within each morphology/stellar mass bin is also shown, revealing that not all bins are equally well represented. Whenever this number is $<10$, the radial profiles are drawn with small points joined by thin dashed lines. The grey dashed line represents the only metallicity value in the model grid (i.~e., the set $\{Z_i\}$) visible given the vertical scale of this plot. Overall, it is observed the well-known behaviour published in most recent works. See the discussion in \S~\ref{sec:metallicity-profiles}.}
\label{fig:mdfs-profiles}
\end{figure*}
The adopted procedure in the literature to build the stellar metallicity radial profiles, is to compute the average value over the azimuthal coordinate, $\theta$, in $N_r$ radial bins. In this study, we take as start point our construction of the MDF in Eq.~\eqref{eq:mdf-definition}. This way, the same radial profiles are equivalent to the first moment of $\text{MDF}(r_k)$ for all $k=1,\ldots,N_r$. In Fig.~\ref{fig:mdfs-profiles} we show the result of this approach, where we follow the binning in morphological class and stellar mass shown in Fig.~\ref{fig:sample-distributions}. Each point in Fig.~\ref{fig:mdfs-profiles} is the result of combining the MDFs within each (morphology and stellar mass) bin following the Eq.~\eqref{eq:mdf-bin}, and then computing the first moment of the resulting MDF in the radial bins $\{r_k\}$. The joining lines hold no physical meaning and serve the sole purpose of guiding the eye. The number of galaxies in each morphological class and stellar mass bin are also shown. Whenever the count of galaxies in these bins is $<10$, the profiles are drawn in small points and the joining lines are dashed and thin.

\update{In the upper left panel of Fig.~\ref{fig:mdfs-profiles}, where all morphologies are blended together, a clear dependency of the slope of the radial profiles on the stellar mass appears: more massive galaxies ($\log{M_\star/\text{M}_{\sun}}>10.5$) seem to have grown their stellar metallicity from the inside towards the outside, consistent with the inside-out scenario \citep[e.~g.,][]{Greene2015, Zheng2017, Ellison2018, Oyarzun2019, Breda2020N, Zibetti2020, Lacerna2020, Santucci2020}}. This trend becomes shallower towards intermediate mass galaxies ($\log{M_\star/\text{M}_{\sun}}\sim10$) and flattens for low mass galaxies ($\log{M_\star/\text{M}_{\sun}}\leq10$). In the case of early-type galaxies (E and S0), intermediate and low mass bins display flatter profiles and even mildly inverted, suggesting a stellar metallicity growth from the outside towards the inside. This result would be consistent with the outside-in scenario \citep[e.~g.,][]{Lin2019}. However, the low number of galaxies in some bins blunt any attempt to draw statistically robust conclusions. For late-type galaxies, the segregation of the profiles with the integrated stellar mass is less clear. It is clear though, that the stellar populations progressively reaches lower stellar metallicities in the central regions $r_k/R_e<1$, from $[Z/\text{H}]>0.0$ (Sa and Sb) through $[Z/\text{H}]\sim0.0$ (Sb, Sbc) towards $[Z/\text{H}]\lesssim0.0$ (Sd), where the stellar metallicity grows outside-in. It is worth to note that the profiles of Sbc and Sc galaxies seem to be a transition between entire inside-out and outside-in stellar metallicity growth, where the inversion of the profiles occur at $r_k/R_e=1.5$~---~$2.1$.

\update{If we focus our attention in the stellar metallicity profiles resulting from well populated morphology and stellar mass bins ($\geq10$ galaxies), the overall trends mirrors previous studies using similar samples \citep{Perez2013, GonzalezDelgado2014a, Goddard2017, Li2018, Parikh2019}, the underlying physics of which was recently reviewed in detail for a much larger sample \citep{Sanchez2020}.} In particular, \cite{GonzalezDelgado2014a} found a correlation between the stellar mass surface density and the stellar metallicity radial structure: more dense regions within galaxies tend to evolve in average stellar metallicity in shorter timescales than less dense ones. Similarly, they reported a correlation between the integrated stellar mass and the overall scale of the stellar metallicity radial structure: the most massive galaxies have steeper metallicity gradients whilst less massive galaxies tend to have flattened profiles and even inverted (mainly for Sd, as shown in Fig.~\ref{fig:mdfs-profiles}). Even though this later trend is weakly manifested in the combined contribution of all morphological types (upper left panel in Fig.~\ref{fig:mdfs-profiles}), the lack of sufficient number statistics in intermediate and low stellar mass bins, for individual morphological classes, render this trend as unreliable for $\log{M_\star/\text{M}_{\sun}}<10$ in our results.

Albeit the radial stellar metallicity profiles have enriched our knowledge of the growth of the stellar content and gas flows (inward or/and outward) in galaxies, they only contain a bit of the whole star formation and chemical enrichment histories. In the following sections we will therefore introduce the resolved stellar metallicity structure of the galaxies in the CALIFA sample described in \S~\ref{sec:sample} using the method explained in \S~\ref{sec:mdf-method}.

\subsection{Global MDF of CALIFA galaxies}\label{sec:mdf-global}

\begin{figure*}
\includegraphics[scale=0.45]{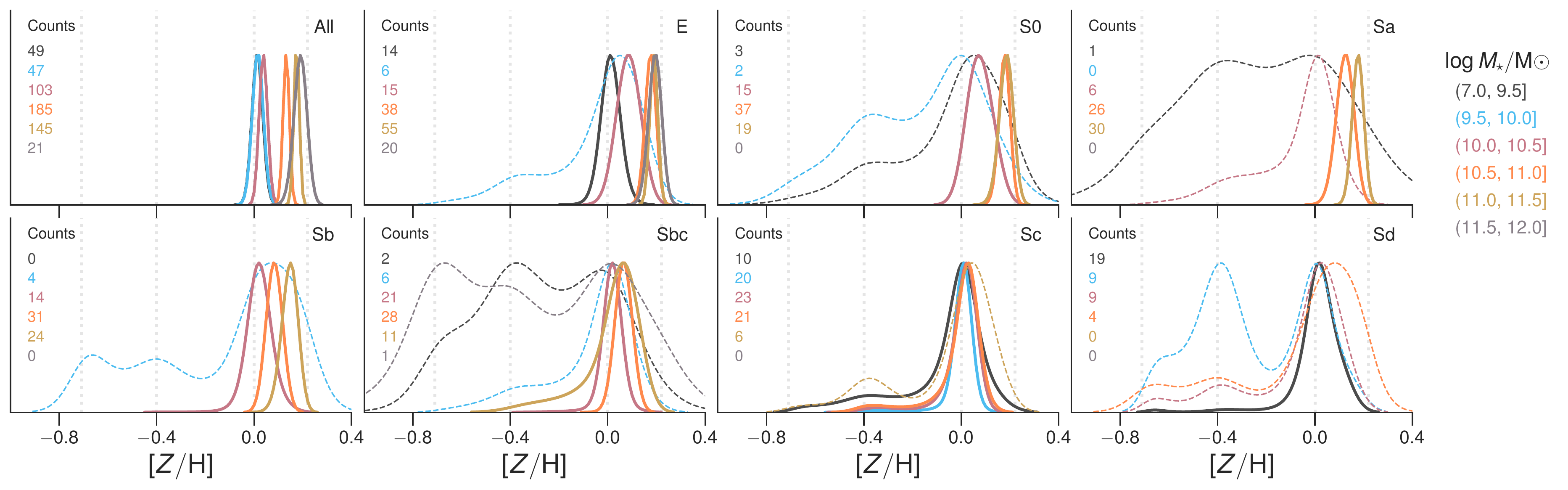}
\caption{The global MDFs for the galaxies in the sample described in \S~\ref{sec:sample} and segregated in the same fashion as in Fig.~\ref{fig:sample-distributions}. As in Fig.~\ref{fig:mdfs-profiles}, the number of galaxies in each morphology/stellar mass bin is also shown in each panel. The MDFs that fall in bins with $<10$ counts are drawn in thin dashed lines. Clearly those MDFs are wider than those in well sampled bins, and are the same producing outlying radial profiles in Fig.~\ref{fig:mdfs-profiles}, specially for E, S0 and Sa morphologies. There is a variety of behaviours in the global MDFs in reliable bins (solid lines). However, altogether for the narrow MDFs, the metallicity profiles (c.~f. Fig.~\ref{fig:mdfs-profiles}) seem to describe the chemical structure of the galaxies, as expected. See \S~\ref{sec:mdf-global} for details.}
\label{fig:mdfs-global}
\end{figure*}
In Fig.~\ref{fig:mdfs-global} we show the global MDFs computed using Eq.~\eqref{eq:mdf-bin} as in the previous section, then we collapse (marginalize) the resulting MDF in each morphological class and stellar mass bin over the radial dimension, using the relation in Eq.~\eqref{eq:mdf-galaxy}. Therefore, the spread in this global MDFs is the propagated uncertainty in the radial profiles in Fig.~\ref{fig:mdfs-profiles}. When the number of galaxies in each morphological class and stellar mass bin is lower than $10$ galaxies, we draw the MDF using dashed thin lines. In the upper left panel in Fig.~\ref{fig:mdfs-global}, where we show the global MDFs in all morphological classes combined, we can see the MDFs are clearly consistent with the radial profiles shown in Fig.~\ref{fig:mdfs-profiles}. These MDFs are narrow enough so that the first moment is a fair representation for the typical stellar metallicity across the galaxies.
However, we warn some MDFs fall in stellar mass bins that are under-sampled across morphological bins as in Fig.~\ref{fig:mdfs-profiles}. This is specially the case of the individual morphologies shown in Fig.~\ref{fig:mdfs-global}, where the three lowest stellar mass bins in the interval $\log{M_\star/\text{M}_{\sun}}=\left(7.0,10.5\right]$, systematically exhibit large-variance MDFs.
At higher (well-represented) stellar mass bins ($\log{M_\star/\text{M}_{\sun}}>10.5$), the peak of the MDFs shifts towards higher stellar metallicities with increasing stellar mass. This trend is consistent with the aforementioned relation between the overall scale of the radial profile and the integrated stellar mass. Both are reminiscent to the stellar mass-metallicity relation \citep[e.~g.,][]{Gallazzi2008, Panter2008, ValeAsari2009}.

\subsection{Resolved MDF of CALIFA galaxies}\label{sec:mdf-resolved}

\begin{figure*}
\begin{flushleft}
\includegraphics[scale=0.45]{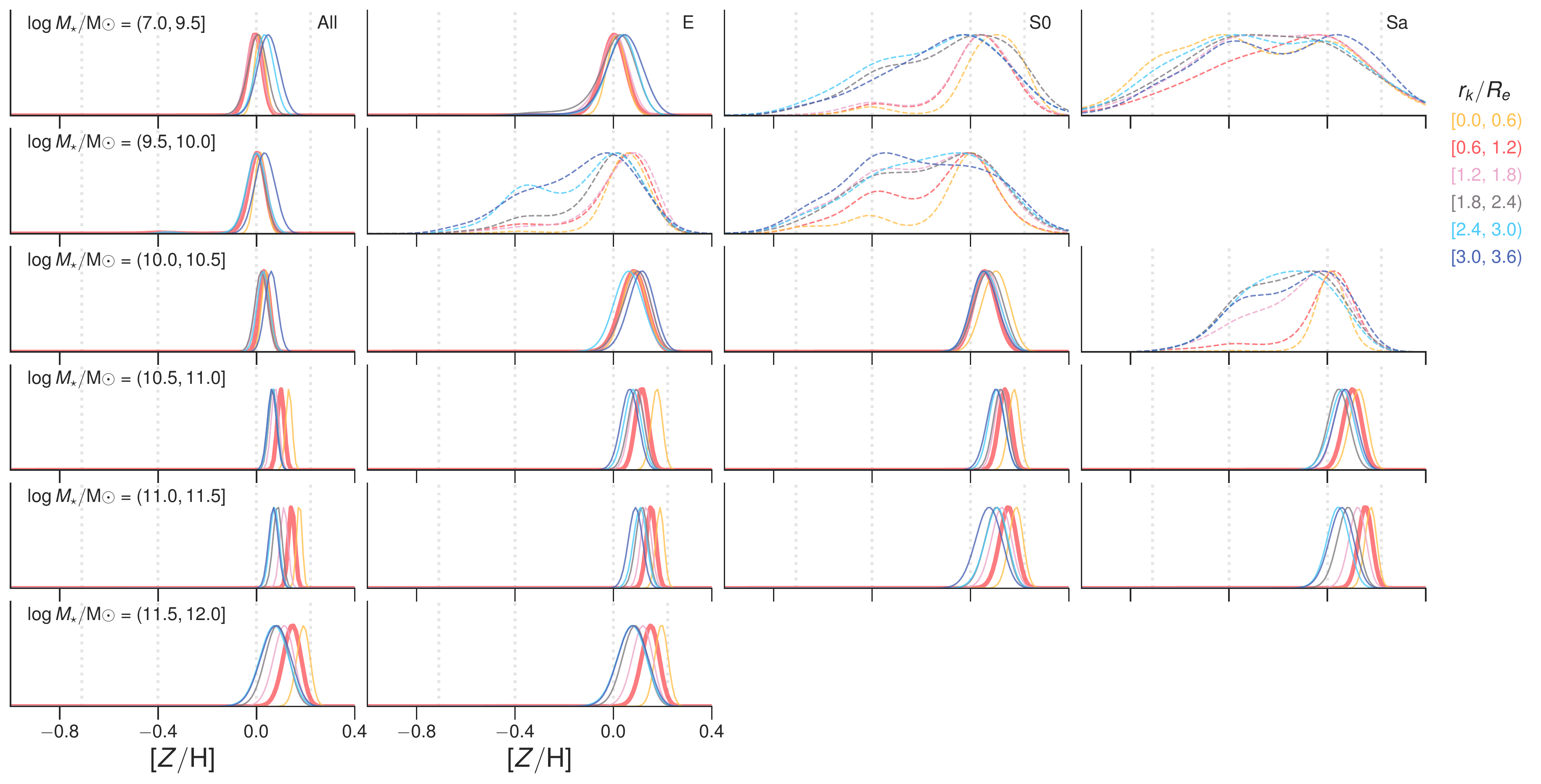}
\end{flushleft}
\caption{The spatially resolved MDFs of galaxies are shown in the same stellar mass bins (rows) as in Fig.~\ref{fig:sample-distributions}. Each column corresponds to different morphological classes: \textit{(i)} all morphologies; \textit{(ii)} ellipticals; \textit{(iii)} S0 and finally \textit{(iv)} Sa. The missing panels agree with the empty bins in Fig.~\ref{fig:mdfs-global}. The radial bin around $1\,R_e$ is highlighted with a thicker red line. The MDFs in under-sampled bins are drawn in thin dashed lines as in Fig.~\ref{fig:mdfs-global}. The stellar mass seems to be a more important driver of the shape of the MDFs than is the morphology. As a function of the stellar mass, the MDFs tend to become more narrow from the less massive galaxies towards high mass counterparts. A secondary trend relates the spread of the MDFs to the radial position: inner regions seem to have higher stellar metallicity than external regions. See the details in \S~\ref{sec:mdf-resolved}.}
\label{fig:mdfs-resolved-early}
\end{figure*}

\begin{figure*}
\begin{flushleft}
\includegraphics[scale=0.45]{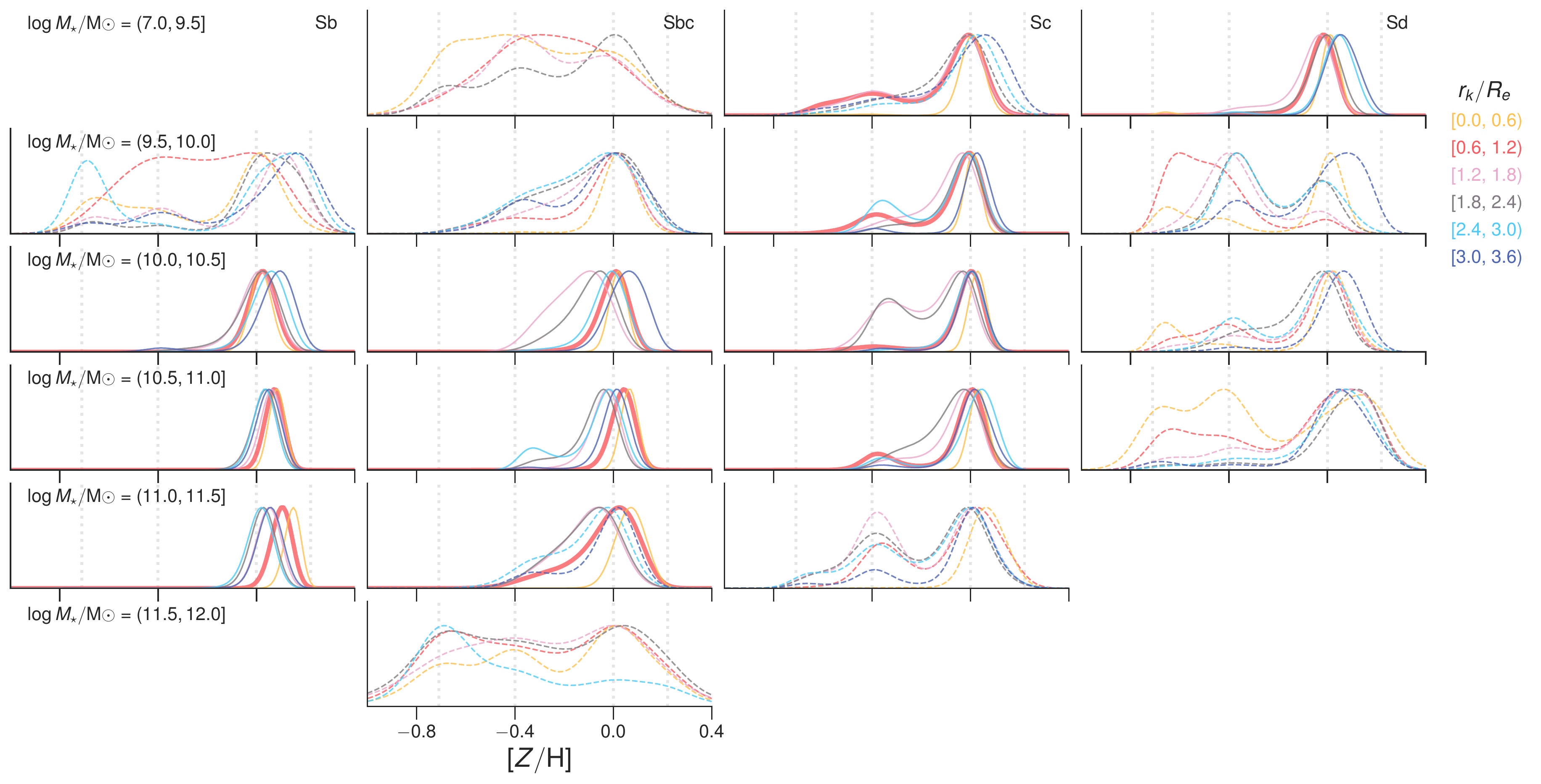}
\end{flushleft}
\caption{
Same as Fig.~\ref{fig:mdfs-resolved-early} but each column corresponds the morphological classes: \textit{(i)} Sb; \textit{(ii)} Sbc; \textit{(iii)} Sc and \textit{(iv)} Sc. As in the global MDFs (Fig.~\ref{fig:mdfs-global}), the under-sampling of the bins has the effect of widening the MDFs. In the case of well-sampled bins, morphological classes Sb and Sbc appear to follow the same trends as the early-type galaxies. However, for morphologies Sc and Sd the stellar mass is not a clear driver in the chemical enrichment. Instead, the MDFs are systematically more spread at all stellar masses with no clear trend. See details in \S~\ref{sec:mdf-resolved}.}
\label{fig:mdfs-resolved-late}
\end{figure*}

In Figs.~\ref{fig:mdfs-resolved-early} and \ref{fig:mdfs-resolved-late} we show the resolved MDFs as a function of both, the morphological class (All morphologies, E, S0 and Sa; columns) and the stellar mass bins (rows) as in Fig.~\ref{fig:sample-distributions}. The radial distance at which the MDFs are computed (color coded) are consistent with the radial bins in Fig.~\ref{fig:mdfs-profiles}. The missing panels represent morphological/stellar mass bins with no galaxies in our sample and whenever the number of galaxies is $<10$ the MDFs are drawn in dashed lines. The radial bin around $1\,R_e$ is highlighted with a thick (red) line. In the left-most column, where all morphological classes are combined, there is a trend with stellar mass in which the MDFs tends to become narrower with increasing values up to $\log{M_\star/\text{M}_{\sun}}\sim11$. In the highest stellar mass bin, however, the MDFs mildly increase their variance. This behaviour holds irrespective of the morphological class (from E to Sa) and the region within the galaxies, if we ignore the under-sampled stellar mass bins (dashed lines). Along with the shift of the MDFs peaks, following the stellar mass-metallicity relation, this result could imply shorter timescales in the chemical enrichment for more massive galaxies \citep[e.~g.][]{GonzalezDelgado2014a}. In this line of thought, a second trend in which the MDFs become narrower towards inner regions of the galaxies, suggests that external regions within the galaxies have delayed ChEHs compared to central ones. This is more evident in S0 --- Sbc morphological classes than in earlier ones \citep[e.~g.,][Camps-Fari\~na, submitted to the MNRAS]{Perez2013, Lin2019}.

The picture drawn from these trends is consistent with that already known from the stellar mass surface density and metallicity profiles \citep[both reviewed in][]{Sanchez2020}. Denser regions imply a higher early SFR and the formation of a large number of massive stars, therefore higher metallicities are reached in shorter timescales \citep{Jerabkova2018}. The radial behaviour of the MDFs becomes less obvious for Sc and Sd galaxies even in all well-represented stellar mass bins. This suggests that there is no clear dependence of the timescale for the chemical enrichment on the radial distance for these type of galaxies. Furthermore, some bimodal features appear in Fig.~\ref{fig:mdfs-resolved-late}. These are commonly found in the Sc morphological class
However, we reckon again that the number of regions in each morphological/stellar mass bin may be part of the reason for the observed variance in these MDFs.

Even though the low number of galaxies is undoubtedly limiting our interpretation of the stellar chemical structure in some morphological/stellar mass bins, Sc and Sd galaxies in stellar mass bins $\log{M_\star/\text{M}_{\sun}}=\left(9.5,11.0\right]$ and $\left(7.0,9.5\right]$, respectively, do show signs of intrinsic broadening (if compared to other morphological classes) likely due to complex chemical mixtures. This result allows us to draw our first important finding: \emph{the stellar metallicity radial profile is not always a good representation of the underlying chemical structure.}
In particular, the observed radial trends of the bimodalities in the Sc morphological class is interesting and may suggest the presence of substructures (common in this type of galaxies) with different SFH timescales, hence different stellar metallicity mixtures. We will elaborate on other physical reasons that can explain the multi-modal behaviour in the MDFs.

\section{Discussion}\label{sec:discussion}

In this section we analyze our main findings in the context of our state-of-the-art knowledge about the evolution of stellar populations within galaxies and their spatially resolved distributions. In \S~\ref{sec:mdf-method} we introduced a novel method for studying the underlying stellar chemical structure of galaxies via fossil record recovery. Even though these fossil record techniques have been extensively used in the literature of extragalatic surveys, it is still worth to note several caveats relevant to this study. First, there is a dependence of the physical properties estimated (stellar mass, stellar age, stellar metallicity) on the model ingredients being used (stellar spectral libraries, evolutionary tracks, etc.), each of which has an associated uncertainty \emph{a priori} unknown in most cases \citep[e.~g.,][]{Walcher2011, Conroy2013}. \update{Second, also unknown are the correlations among the physical properties being recovered. Such is the case of the outshining effect of young stellar populations \citep[$1\,$Gyr, e.~g.,][]{Maraston2010} and other sources of degeneracies between the stellar mass, age metallicity and dust extinction \citep[e.~g.,][]{Conroy2013}. Third, the chosen minimization algorithms across the literature have shown disagreements when tested against mock data sets \citep[e.~g.,][]{Guidi2018, Ibarra-Medel2019}.} Such discrepancies eventually translate into part of the variance observed in the stellar properties radial profiles, specially in the stellar metallicity \citep[e.~g.,][]{Sanchez-Blazquez2014, GonzalezDelgado2015, Goddard2017, Zhuang2019, Oyarzun2019}.

Naturally the uncertainties propagating from the model ingredients and spectral fitting methods are likely playing an important r\^ole. Possibly the most insidious of such uncertainties comes from the fact that most SSP models (including the ones adopted in this study) can only predict solar abundances of heavy species. Therefore complex patterns such as the $\alpha$-enhancement found in stellar populations that evolved in a short timescale, can not be accurately estimated. In general, we expect the resulting inadequacies to be more relevant in high surface stellar density regions within late-type galaxies (e.~g. bulges) and in early-type galaxies. On the other hand, the difference usually found for star-forming late-type galaxies, between the light- and the mass-weighted stellar properties (even when estimated using the same procedures), are telling a more physically relevant story: the spread of the underlying distribution of such properties. Therefore, it is interesting, in the light of our results that some of such discrepancies, at least in the stellar chemical structure of galaxies, can in fact be justified by the spread of the MDFs itself. Hence, by the inner workings defining the SFH and ChEH of these galaxies.

In \S~\ref{sec:metallicity-profiles}, if we focus only on the statistically reliable profiles with $\geq10$ galaxies, the trends already known from the radial profiles of the stellar metallicity emerged. Notwithstanding, in \S~\ref{sec:mdf-global}, some of the intricacy of the chemical structure of galaxies, specially for late-type ones, also emerged (large variance, multi-modal and asymmetrical behaviours in the MDFs). Several physical reasons may explain the observed features in the MDFs: \textit{(i)} galaxies in such morphological class and stellar mass bins have intrinsically a more complex stellar metallicity mixture; \textit{(ii)} galaxies have also a SFH with large spread in time, therefore the ChEH has occurred in a larger timescale than in galaxies within other stellar mass and morphological class bins; \textit{(iii)} there are two or more distinctive substructures (projected along the line-of-sight) shaping those distributions; or, most likely, \textit{(iv)} a combination of the above.

\begin{figure*}
\includegraphics[scale=0.45]{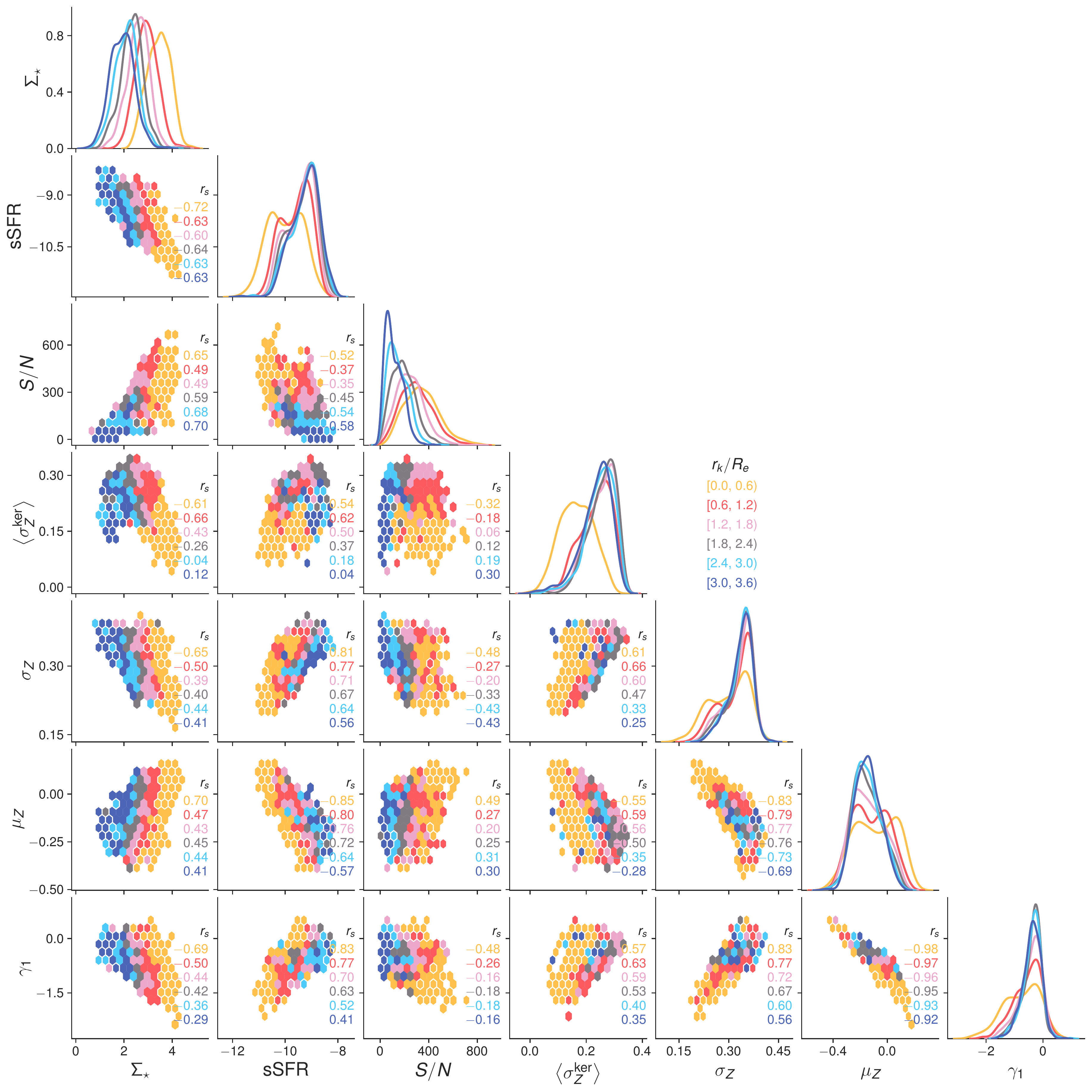}
\caption{
The distribution of a set of resolved physical/observational properties ($\Sigma_\star$, sSFR, $S/N$) along the parameters defining the shape of the MDFs ($\left<\sigma_Z^\text{ker}\right>$, $\sigma_Z$, $\mu_Z$, $\gamma_1$), are shown for the galaxies in the explored sample. Each hexagonal bin is color-coded according to the most frequent galactocentric distance therein and outlined with a thick line if dominated by multi-modal MDFs. Only bins with $>3$ number of regions are shown. As a measure of the strength of the relationship between each property, the Spearman correlation coefficient ($r_s$) within each radial bin, is also shown. On top of each column, the KDE of each parameter is shown following the same color coding. Several strong correlations ($|r_s|>0.6$) can be appreciated, among which those relating physical properties to MDFs moments are the most relevant. See the text in \S~\ref{sec:discussion} for details.}
\label{fig:mdfs-correlations}
\end{figure*}
To test for those hypotheses, in Fig.~\ref{fig:mdfs-correlations} we show the distributions of a set of resolved physical/observational properties: the stellar mass surface density ($\Sigma_\star$), the specific star-formation rate ($\text{sSFR} \equiv \Sigma_\text{SFR} / \Sigma_\star$, where $\Sigma_\text{SFR}$ is the resolved SFR averaged over the last $100\,$Myr) and the signal-to-noise ratio ($S/N$), against the distribution of several moments of the resolved MDFs: the average kernel standard deviation ($\left<\sigma_Z^\text{ker}\right> \equiv \sum_{j}\,w_j\sigma_{Z,j} / \sum_{j}\,w_j$), the MDF mean stellar metallicity ($\mu_Z$, defined in Eq.~\ref{eq:kernel-mu}), the MDF standard deviation ($\sigma_Z$, defined in Eq.~\ref{eq:kernel-sigma}) and the skewness ($\gamma_1$). The strength of the correlation between the different parameters, at different radial distances, is estimated based on the Spearman correlation coefficient ($r_s$) calculated over the individual samples in each radial bin. It is worth to note that several correlations between physical properties and the MDFs moments appear. We try to explain the variety of behaviours in the shapes of the MDFs seen in previous sections, in the framework of these correlations and its physical implications.

In Fig.~\ref{fig:mdfs-correlations}, the correlations with the radial distance are among the weakest, except for the stellar mass surface density. This strong trend shows that at small (large) radial distance, high-density (low) regions are systematically found. This property is also correlated with the $S/N$ ($r_s=-0.61$).
More relevant, however, are the weakness of correlations between the $S/N$ and other properties in this plot ($|r_s|\sim0.30$) and the correlations between $\Sigma_\star$ and the moments of the MDFs: the average kernel standard deviation ($r_s=-0.45$) and, the MDF standard deviation ($-0.42$), the mean ($-0.40$) and the skewness ($-0.43$). These correlations show an increasing strength towards central regions of the galaxies ($r_s\sim0.65$). The trends with $S/N$ (or lack thereof) suggest that the observations quality is not biasing our results, despite the fact that there is indeed a radial gradient in $S/N$.


The most tight correlations are found among the MDFs moments. The MDFs mean (i.~e., the average stellar metallicity) tend to be higher when its standard deviation is low ($r_s=-0.81$) and the skewness strongly tends towards negative values ($r_s=-0.94$). Conversely, when the MDFs have lower mean (i.~e., low stellar metallicity regions), the standard deviation is higher and the skewness is consistent with symmetric distributions ($\gamma_1\sim0\,$dex). The fact that these strong correlations exist imply a common origin in the physics setting the SFHs and ChEHs in galaxies and, by extension, the MDFs. On the other hand, the trends between the average kernel standard deviation distributions and the radial distance shows a distinctive behaviour in the inner-most radial bin: the average kernels in these regions of the galaxies are narrow ($\left<\sigma_Z^\text{ker}\right>\leq0.1\,$dex), while in outer regions a wide range of standard deviations can be seen, but mostly biased towards higher values ($>0.2\,$dex). Perhaps more interesting is the fact that, despite such behaviour in the spread of the average kernel, the resulting MDFs have a wide range of standard deviations. These apparent discrepancy is noteworthy. Albeit these two moments are both integrated along the SFH timescale, they differ on how strongly biased they are. $\left<\sigma_Z^\text{ker}\right>$ is more biased towards the spread of the kernel that contributes the most to the resulting MDF, whereas $\sigma_Z$ is simply the spread of the MDF itself. Put in physical terms, both moments measure the ChEH timescale, however $\left<\sigma_Z^\text{ker}\right>$ is more sensitive to shorter timescales than $\sigma_Z$. For this reason we use the former as a proxy for the intrinsic stellar metallicity mixture at a given radial distance within the galaxies.

\begin{figure}
\includegraphics[scale=0.45]{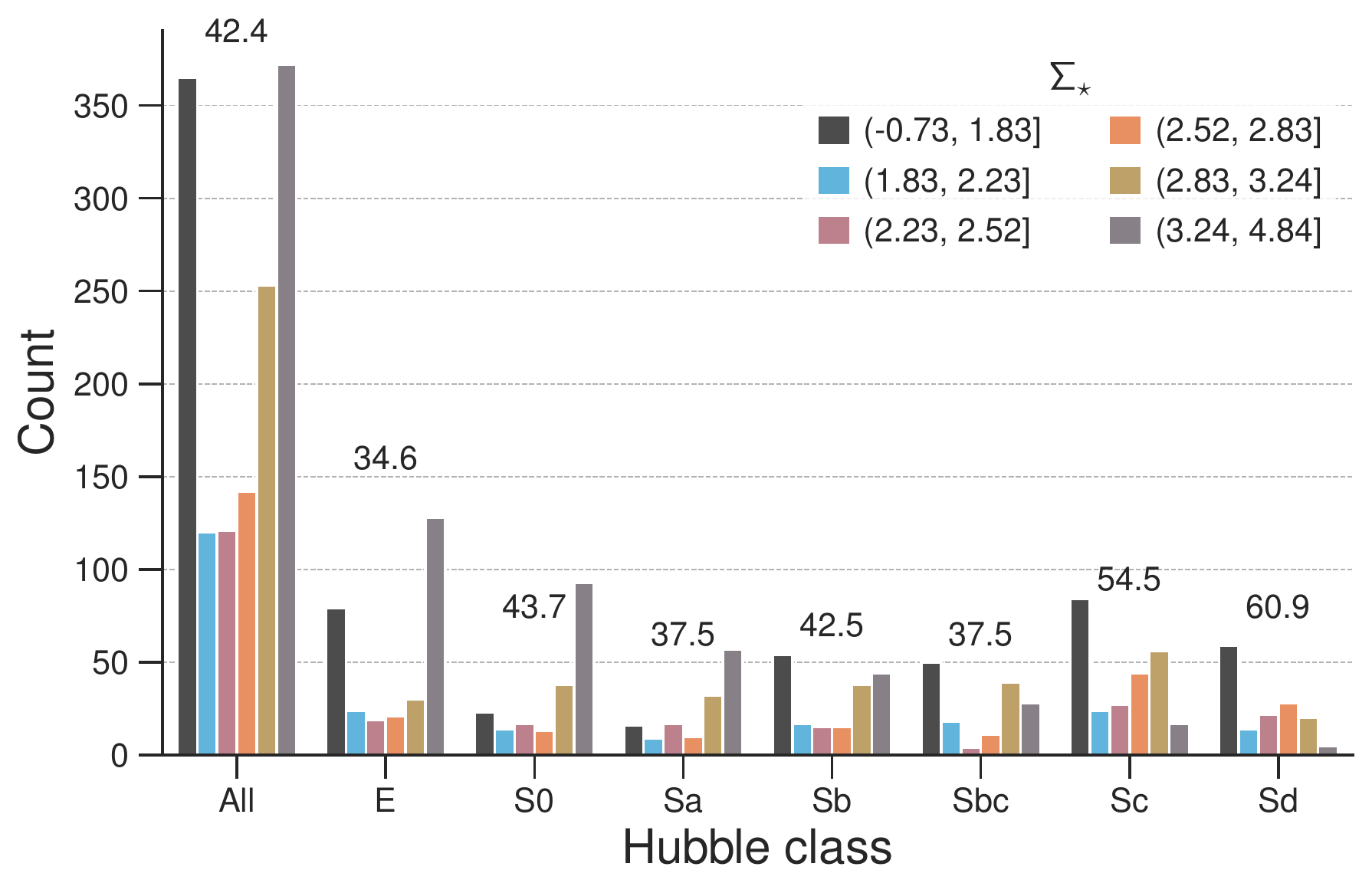}
\caption{Similar to Fig.~\ref{fig:sample-distributions},
the number distribution of radial bins in galaxies with multi-modal MDFs ($\sim1300$) segregated by morphology and $\Sigma_\star$. The numbers on top of each histogram show the percentage of regions with multi-modal MDFs relative to the total number of regions ($\sim3200$), for each morphological class. The distributions show in general two peaks at the highest and lowest $\Sigma_\star$ irrespective of the morphology.}
\label{fig:mdfs-multimode-distributions}
\end{figure}
In Fig.~\ref{fig:mdfs-multimode-distributions} we show the number distribution of the radial bins containing multi-modal MDFs, as a function of the stellar mass surface density thereof and the morphological class of the hosting galaxy. The overall contribution of regions with these MDFs is low, surpassing $42\,$per cent. The distribution containing all the morphological classes reveals two peaks, one at low density regions ($\Sigma_\star\sim1.1\,$dex) and the other at high density regions ($\Sigma_\star>3.5\,$dex). This trend is present at all morphological classes, except for the S0, Sc and Sd. One possible explanation for this outlying behaviour in these morphologies is the small number of galaxies in our sample (c.~f. Fig.~\ref{fig:sample-distributions}). The frequency of the multi-modal MDFs in low density regions seems to increase towards late-type galaxies (from S0 to Sc) in absolute terms, i.~e., with respect to the total number of regions. Even though the fraction of these regions with complex MDFs is relatively low, the frequency of galaxies with multi-modal stellar metallicity distributions at any radial bin, per morphological class is generally over $90\,$per cent. Certainly, in late-type galaxies these MDFs are more common ($>95\,$per cent).

\subsection{SFH and ChEH timescales}

By construction, our method for estimating the MDF involves the integration through the SFH timescale. The more distributed are the weights $w_{ij}$ along the $j$ dimension, the longer has taken the stellar mass assembly to build up to the present state (see Eq.~\ref{eq:kernel-mu}). Consequently, if the weights $w_{ij}$ are fairly shared along the kernels of complex or in-homogeneous metallicity mixtures, the resulting MDF will have a large standard deviation, with few features. This is the case of the most outer radial bins in morphological classes Sbc and Sc in Fig.~\ref{fig:mdfs-resolved-late}. On the other hand, if the weights are distributed along two or more well defined (small variance) and separated kernels, this would translate into the observed multi-modal behaviour. This is the case in the most inner radial bins, where is likely the disc and the bulge are both shaping the MDFs. Nevertheless, the existence of multi-modal MDFs in outer regions of the galaxies is not clear and can be due to substructures (bar, spiral arms) extending in these regions of the galaxies ($r_k/R_e>1.8$) and/or radial migrations (see next section). In both cases the signature in the SFH and ChEHs would be an extended timescale with respect to inner and homogeneous regions (in the $Z$ sense).

To measure the spread in time of the SFH and ChEH across the galaxies extent, we use the resolved sSFR defined as the fraction of the total stellar mass that has been assembled during the last $100\,$Myr within each radial bin. The regions that have assembled their stellar mass later in time have higher values of sSFR than those that have assembled early have lower values of sSFR \citep{LopezFernandez2018}. The correlation between this parameter and the standard deviation of the MDFs ($r_s\sim0.71$ in Fig.~\ref{fig:mdfs-correlations}) suggests that regions that are still assembling their stellar content (higher sSFR) result in wide MDFs. On the other hand, those regions in which the stellar content grew in shorter timescales (lower sSFR) yield narrower MDFs. It is worth to note the relative strength of the correlation between the average kernel standard deviation ($\left<\sigma_Z^\text{ker}\right>$) and the MDF spread ($\sigma_Z$) across the radial extent of galaxies. In the outer regions ($r_k/R_e>0.6$) the correlation coefficient $r_s\sim0.40$ in contrast to what is seen in the most inner regions ($r_k/R_e<0.6$) with $r_s=0.61$. This hints that the SFH and ChEH timescales are not the relevant factor in shaping the the metallicity distributions in the outskirts of the discs, although they are indeed relevant to define the average stellar metallicity \citep{GonzalezDelgado2014a}.

\subsection{Complex stellar metallicity mixture}

There are two levels in which we can measure the complexity of the stellar metallicity mixture in our MDF estimates. One is at the kernel level, which is provided by the standard deviation in Eq.~\eqref{eq:kernel-sigma} and is determined by the share of the weights $w_{ij}$ along the $i$ dimension. The more shared the weights along $i$, the more complex is the stellar metallicity mixture at a given stellar age and radial distance in the galaxy. The second is at the level of the SFH, which is related to the spread of the weights $w_{ij}$ along $j$. This is essentially the SFH timescale. We discussed the later in the previous section. As for the former, the average kernel standard deviation is a direct measure of the intrinsic metallicity mixture.

Certainly, the negative correlation between the average kernel standard deviation and the stellar mass surface density ($r_s\sim-0.5$ in Fig.~\ref{fig:mdfs-correlations}) indicates that the mixture of stellar metallicities at a certain time of the SFH and radial distance, becomes more homogeneous at denser and inner-more regions. Likewise, the negative correlation between the stellar mass surface density and the sSFR suggest that these regions have grown their stellar mass in shorter timescales ($>90\,$per cent of the mass assembled before the last $100\,$Myr), than their outer counterparts \citep[local downsizing,][]{Perez2013}. Consequently, the stellar metallicity distributions arising from these central galactic regions seems to be shaped by well-defined (narrow) kernels.

The fact that most of the multi-modal MDFs correspond to high-density regions ($r_k/R_e<1$) suggests that they have arose from an early fast chemical enrichment (producing the high-metallicity peak) followed by: \textit{(i)} a late star-formation event from metal-poor gas;  \textit{(ii)} the coalescence (in the line-of-sight) of substructures with intrinsically distinct SFHs and ChEHs, such as the bulge and the disc; \textit{(iii)} the mixture of distinct stellar populations due to the (2D) projection of triaxial substructures, such as the bulge; and/or \textit{(iv)} the radial migration of outer metal-poor populations towards more inner regions. These scenarios are not mutually exclusive, and disentangling or even detecting their imprints left in the SFHs and ChEHs, is an ongoing endeavour. For instance, \cite{Breda2018} found that the bulge in late-type galaxies may have formed through a combination of a fast stellar mass growth in early stages ($\sim3\,$Gyr since the SFH onset) plus continuous star formation and (inward) radial migration of star-forming clumps, consistent with our results \citep[see also][and references therein]{Breda2020N}.

On the other hand, low density regions ($r_k/R_e>1$) correspond to relatively homogeneous (early-type/bulge-dominated galaxies in Fig.~\ref{fig:mdfs-resolved-early}) or in-homogeneous (late-type galaxies in Fig.~\ref{fig:mdfs-resolved-late}) mixtures of stellar metallicities. The later can be attributed to either episodic star-formation events and/or radial migrations. Further to this, the r\^ole of azimuthal structures in the gas-phase and stellar metallicity has been proposed to explain the observed variance in the radial profiles of late-type galaxies \citep[e.~g.,][]{Ruiz-Lara2017, Sanchez-Menguiano2018}. In particular, \cite{Sanchez-Menguiano2016} found evidence of ongoing transport of material due to the spiral structure of NGC~6754. According to their results, metal-rich (poor) gas travels from the inner (outer) disc towards the outer (inner) parts of the disc in this particular galaxy. Their findings are in agreement with the hypothesis of transient spiral arms, whereby the long-lasting radial transport of gas would turn into the observed in-homogeneous mixtures of stellar populations \citep[see also][]{Ruiz-Lara2017a, Zhuang2019, Dale2020}.

The link between the kinematics and the MDF has been investigated in studies of the MW and MW-like (simulated) galaxies. \cite{Hayden2015} made a reconstruction of the MW disc MDF at several radial and height bins. They found that in the thin disc ($|z|<0.5\,$kpc) the MDF across the radial extent shows an inversion of the skewness from negative values at the inner disc ($R<7\,$kpc) to positive values in the outskirts of the disc. They showed through a set of simulations that radial migration not only explains the skewness inversion but the skewness itself. In the same line of thought, \cite{Martinez-Medina2017} implemented chemo-dynamical simulations of MW-like galaxies and quantified the r\^ole of the substructures (e.~g., bar, spiral arms) and their kinematics in shaping the MDF across the radial extent of galaxies. They demonstrated that the MDF may encode structural information of the spiral arms, provided they are the main responsible for radial migrations of stellar populations \citep[see also][]{Kubryk2013, Minchev2014}.

Altogether, the observed spread in the MDFs (ignoring under-sampled bins) unambiguously corresponds to regions of low surface mass density (frequent in outskirts of discs), or galaxies with low surface density across all their optical extent. These regions are further characterized by an intrinsic in-homogeneous stellar metallicity mixture, consistent with radial migration of gas clumps or stellar populations. On the other hand, even when the average kernel of a given MDF shows no signature of a complex mixture of stellar populations, the resulting MDF can in fact reflect it. This behaviour explains the existence of multi-modal MDFs, whereby well-defined narrow kernels with different peak locations strongly contribute in shaping the stellar metallicity distribution. These MDFs are commonly found in the central regions of late-type galaxies (c.~f. Fig.~\ref{fig:mdfs-correlations}), but are fairly common in regions with optical substructures (e.~g., bulge, bars, spiral arms). Albeit we have restricted the physical origin of the variety of shapes in the MDFs, a thorough examination on their relationship with the parameters of the optical and kinematics structures within galaxies is in order. However such endeavour is beyond the scope of the current study and therefore will be deferred to the upcoming research efforts.

\section{Summary and Conclusions}\label{sec:conclusions}

We have introduced a novel method to attempt resolving the metallicity structure of galaxies observed in spatially resolved surveys. It consists in a reconstruction of the underlying distribution using a KDE-like approach. Using this method, we were able to reproduce the known trends in the radial profiles as a function of the morphology and the stellar mass, using a sample of $550$ almost face-on ($i<70\,$deg) galaxies across all morphological classes from the CALIFA survey. Furthermore, we found that for those galaxies (and regions within) where the results in the literature tend to show large variance (e.~g., outskirts of late-type galaxies) coincide with regions where the chemical structure is dominated by a complex composition of stellar populations. Our main results can be summarized as follow:
\begin{description}
\item[\textit{(i)}] The stellar metallicity structure revealed by the radial profiles can be biased if the underlying distribution is dominated by a in-homogeneous metallicity mixture of stellar populations. This is the case of the stellar metallicity in the outskirts of the optical extent of galaxies, as well as most of the optical extent of late-type galaxies. However, this is also true for the inner regions where the stellar populations in the bulge and the disc coexist. In such cases, the radial profile (first moment of the MDF) will hardly convey the true nature of the chemical structure. Hence, to study only the stellar metallicity profiles will lead to incomplete and misleading interpretations of the ChEH of galaxies.
\item[\textit{(ii)}] Overall early-type (E and S0) galaxies display a well-defined metallicity structure, rarely showing large variance or multi-modal behaviours in central regions. The skewness is generally negative, which means an asymmetrical distribution biased towards super-solar stellar metallicities. On the other hand, late-type (Sa~---~Sd) galaxies manifest their complex mixture of stellar populations in their chemical structure in the form of multi-modal, large variance and/or nearly symmetrical MDFs. The dependence of the MDF shape on the stellar mass of the galaxies is such that less massive galaxies ($\log{M_\star/\text{M}_{\sun}}<10$) have broader MDFs, whilst more massive galaxies ($\log{M_\star/\text{M}_{\sun}}>10$) have more narrow ones. However, given the small number of galaxies in some morphological classes and stellar mass bins, the relative r\^ole of these factors in shaping the MDF is still unclear.
\item[\textit{(iii)}] The physical origin of these features in the stellar metallicity structures are directly linked to the SFH and ChEH timescales. On one hand, the large variance mono-modal MDFs can be attributed to SFHs with large timescales, allowing for a vast metallicity mixture in the present. On the other hand, in multi-modal MDFs one super-solar metallicity peak is associated to a fast chemical enrichment in the early stages of the SFH, while the one or two more metallicity-poor peaks are consistent to: \textit{(a)} a posterior star-formation event due to gas inflow, \textit{(b)} the coalescence of substructures with different SFHs, \textit{(c)} the 2D-projection of triaxial substructures, and/or \textit{(d)} radial migrations.
\end{description}

In summary we have attempted the first exploration, to our knowledge, of the shape of the MDF across the Hubble sequence of galaxies outside our local group. Using this pioneering methodology we are able to present the main patterns of the MDF for different masses and morphologies. With this results we opened a new view to study the chemical evolution in galaxies that should be confronted with simulations and adopted to constrain the evolution of individual galaxies.

\section*{Acknowledgements}

We thank CONACYT FC-2016-01-1916 and CB-285080 projects and PAPIIT IN100519 project for support on this study. AMN specially thanks to Carlos L\'opez-Cob\'a who selflessly shared his calculation of isophotal parameters for the sample used in this paper. AMN also thanks to Jorge Barrera-Ballesteros for useful comments that improved an early version of this paper. AMN and EADL thank all Stellar Population Synthesis and Chemical Evolution group of the Instituto de Astronom\'ia - UNAM for the help through the by-eye morphological classification. L.G. was funded by the European Union's Horizon 2020 research and innovation programme under the Marie Sk\l{}odowska-Curie grant agreement No. 839090. R.G.B. acknowledges support from the State Agency for Research of the Spanish MCIU through the ``Center of Excellence Severo Ochoa'' award to the Instituto de Astrof\'isica de Andaluc\'ia (SEV-2017-0709).




\bibliographystyle{mnras}
\bibliography{califa-mdfs} 




\appendix

\section{Probability algebra for MDFs}\label{app:probabilities}

In Eq.~\eqref{eq:mdf-theoretical} we defined the MDF as the probability distribution function conditional on the galactocentric distance $r$, $P\left(Z\,\middle|\,r\right)$. In this section, in order to justify Eqs.~\eqref{eq:mdf-galaxy} and \eqref{eq:mdf-bin}, we establish the mathematical framework for combining the MDFs following the rules of probabilities, namely:
\begin{equation}\label{eq:sum-rule}
\int\,P\left(Z\,\middle|\,r\right)\,\text{d}Z = 1,
\end{equation}
and
\begin{equation}\label{eq:pro-rule}
P\left(Z_A, Z_B\,\middle|\,r\right) = P\left(Z_A\,\middle|Z_B,r\,\right)\times P\left(Z_B\,\middle|\,r\right),
\end{equation}
where $P\left(Z_A, Z_B\,\middle|\,r\right)$ is the probability of observing stellar populations in galaxies $A$ and $B$ having metallicities $Z_A$ and $Z_B$, $P\left(Z_A\,\middle|Z_B,r\,\right)$ is the probability of observing stellar populations characterized by $Z_A$ in galaxy $A$, given that we observed populations with $Z_B$ in galaxy $B$. We note that we implicitly allow for $Z_A=Z_B$ even though such populations occur in different galaxies, i.~e., $Z_A$ and $Z_B$ are still considered different events. If we assume galaxies $A$ and $B$ to be isolated, then such events are independent and the conditional probability becomes $P\left(Z_A\,\middle|\,Z_B,r\right) = P\left(Z_A\,\middle|\,r\right)$. We can generalize this relation to an arbitrary number of galaxies $N$, as follows:
\begin{equation}\label{eq:pro-general}
P\left(Z\,\middle|\,r\right) \propto \prod_{m=1}^{N}\,P\left(Z_m\,\middle|\,r\right),
\end{equation}
which is equivalent to Eq.~\eqref{eq:mdf-bin} and where the proportionality factor ensures that the summation rule in Eq.~\eqref{eq:sum-rule} holds truth. 

The $n$-th moment can be calculated from Eq.~\eqref{eq:pro-general} as:
\begin{equation}\label{eq:nth-moment}
\mu_Z^n(r) \propto \int (Z - \mu_0)^nP\left(Z\,\middle|\,r\right)\text{d}Z,
\end{equation}
where the radial profile is simply the first  moment. We used the formalism in Eq.~\eqref{eq:pro-general} combined with the first moment given by Eq.~(\ref{eq:nth-moment}, with $n=1$ and $\mu_0=0$) to compute the radial profiles in Fig.~\ref{fig:mdfs-profiles}.

In \S~\ref{sec:results} we also defined the \emph{integrated} MDF across a given galaxy extent, as the marginalization integral
\begin{equation}\label{eq:mdf-marginalized}
P(Z) \propto \int P\left(Z\,\middle|\,r\right)\text{d}r,
\end{equation}
which is equivalent to Eq.~\eqref{eq:mdf-galaxy}. We combined this formalism with the given in Eq.~\eqref{eq:pro-general} to compute the global MDFs in Fig.~\ref{fig:mdfs-global}, while in Figs.~\ref{fig:mdfs-resolved-early} and \ref{fig:mdfs-resolved-late} we only used Eq.~\eqref{eq:pro-general} to compute the resolved MDFs.

\section{Method reliability}\label{app:reliability}

In \S~\ref{sec:mdf-method} we introduced a KDE-like method for retrieving the stellar MDF of IFS data sets. In this section we randomly select a subset of $5$ galaxies from our sample to test for the effect of the bandwidth, $h$, in Eq.~\eqref{eq:kde-definition} on the shape of our MDF estimations. Additionally, we present tests on toy models to assess the reliability of such method against the shape of the SFH and ChEH and the signal-to-noise ratio.

\subsection{Effect of the bandwidth}\label{app:bandwidth}

\begin{figure*}
\includegraphics[scale=0.45]{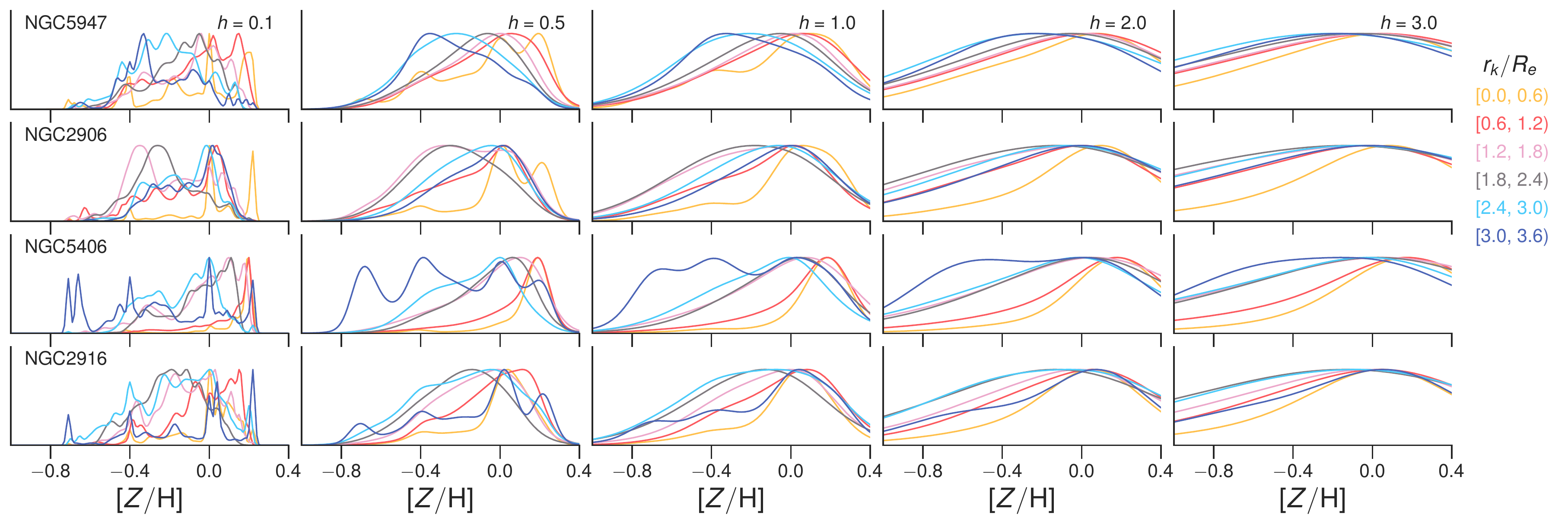}
\caption{The effect of the bandwidth, $h$, is shown for four galaxies in the sample studied in this work: NGC~$5947$, NGC~$2906$, NGC~$5406$ and NGC~$2916$ (rows) and $h=0.1$, $0.5$, $1.0$, $2.0$ and $3.0$ (columns). As expected, for small values of $h$ the variance in the stellar metallicity at each point in time, $t_j$, dominates the shape of the resulting MDFs. For larger values of $h$, the MDFs exhibit less statistical noise from the ChEH sampling, but instead become more biased.}
\label{fig:mdf-bandwidth-test}
\end{figure*}
By definition the KDE has one free parameter: the bandwidth $h$. It serves the purpose of modulate the spread of the chosen kernel with the objective of reducing the statistical variance in the resulting PDF due to poor sampling of the data. A robust method to constrain $h$ is by using a cross-validation method \citep[e.~g., \S~6.1 in][]{Ivezic2019}. However our method is not exactly a KDE (therefore KDE-like) since we restricted the variance of the Gaussian kernel to be the standard deviation of the stellar metallicity mixture $\{Z_i\}$ at a particular instant $t_j$ and radial distance $r_k$ (c.~f., Eq.~\ref{eq:kernel-sigma}). In Fig.~\ref{fig:mdf-bandwidth-test} we show how the shape of the MDF changes as a function of several values of $h$ around the assumed value in this work ($h=1.0$) for a subset of galaxies in our sample. Essentially, the $h$ allows for a trade-off between the variance of the ChEH at each $t_j$ (Eq.~\ref{eq:kernel-sigma}), for $h<1.0$ and, and the bias (Eq.~\ref{eq:kernel-mu}), for values $h>1.0$. In our reconstruction of the MDF, we are interested in $h=1.0$, since (by design) it preserves the physical information given by the spread of the intrinsic metallicity distribution (c.~f., Fig.~\ref{fig:mdf-demonstration} and Eq.~\ref{eq:mdf-definition}). We note that in some cases $h=0.5\,$---$\,2.0$ would provide similar results to the ones discussed in this work.

\subsection{Toy models tests}\label{app:toy-models}

\begin{figure*}
\includegraphics[scale=0.5]{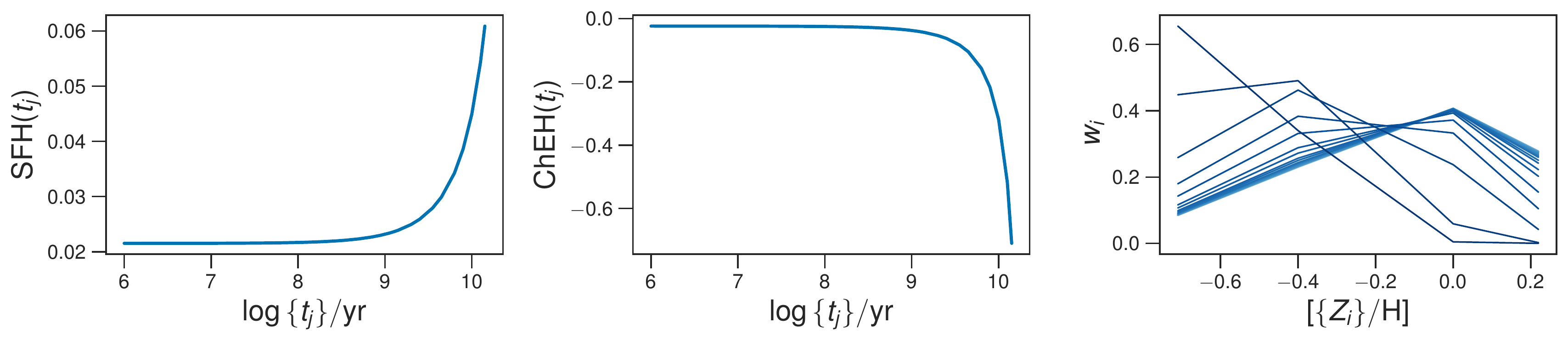}
\includegraphics[scale=0.5]{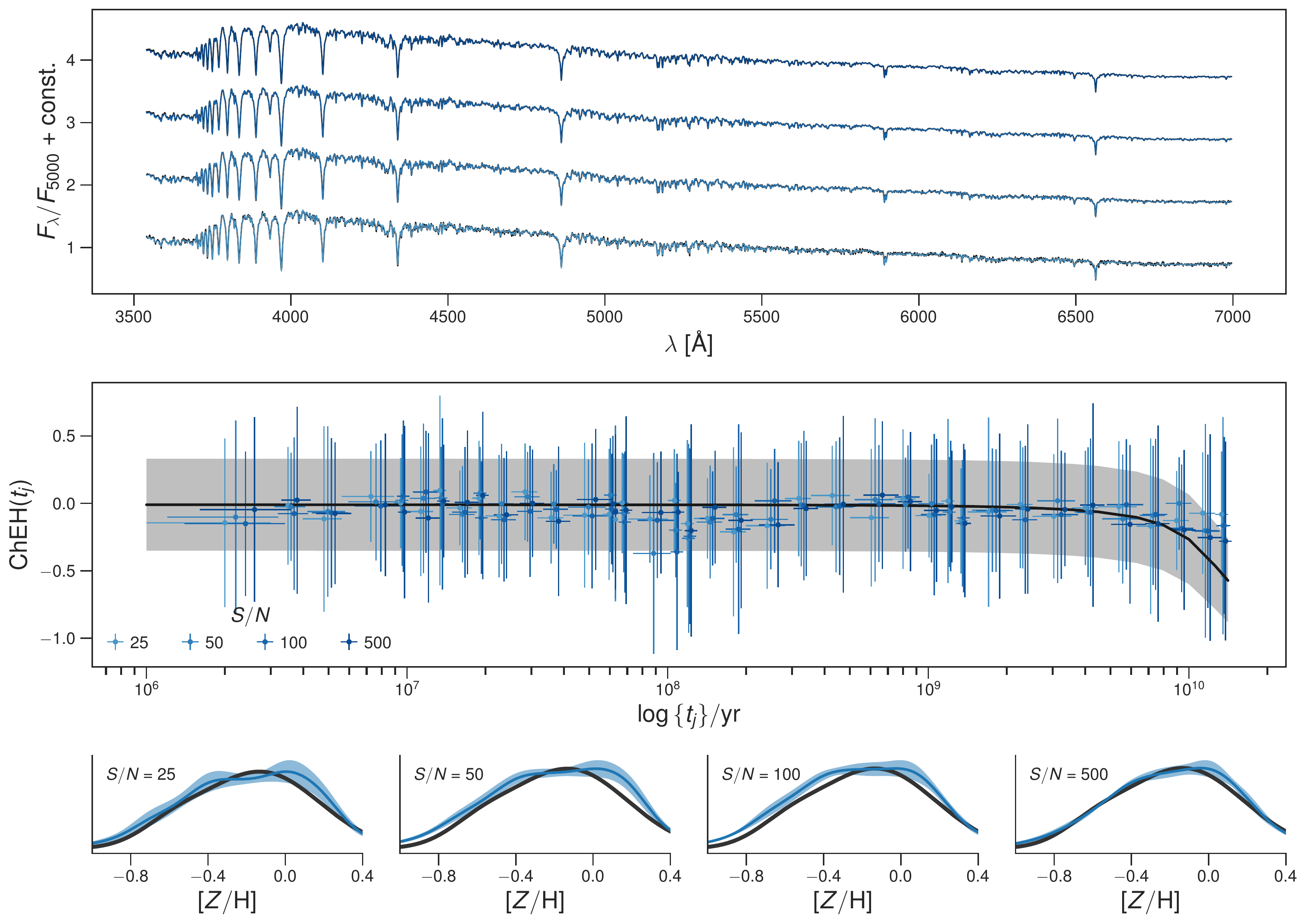}
\caption{
An example of the toy simulations designed to test the MDF method presented in this work. From top to bottom rows: \textit{(i)} the SFH (left), the ChEH (middle) \update{and the distribution of $w_{ij}$ at different ages as a function of metallicity} (right); \textit{(ii)} the input spectra (black), best fitting spectra (blue) and the $1\,\sigma$ region (grey) for the $S/N=25,50,100,500$, from bottom to top; \textit{(iii)} the comparison between the input ChEH (black), along with its intrinsic $1\,\sigma$ spread (grey, c.~f. Eq.~\ref{eq:sim-kernel}) and the retrieved ChEH (blue), where the horizontal bars represent the width of the age bins and the vertical bars are the $1\,\sigma$ range around the average ChEH computed across the $10$ noise realisations; \textit{(iv)} the comparison between the input MDF (black) and the resulting MDFs along with the $1\,\sigma$ region (blue), for each $S/N$ value. The precision and accuracy of the recovered MDF improves with increasing $S/N$, as expected.}
\label{fig:sim-showcase}
\end{figure*}
In this section we simulate SFH and ChEHs using simple $\tau$-models. For the SFR at each time step in the $\{t_j\}$ grid, we assume the functional shape
\begin{equation}\label{eq:sim-sfh}
\text{SFH}(t) \equiv \exp{\left(-t/\tau\right)},
\end{equation}
where $t$ is the time since the onset of the star formation. For the chemical enrichment we assume the shape
\begin{equation}\label{eq:sim-cheh}
\text{ChEH}(t) \equiv (Z_\text{max} - Z_\text{min})\tanh{\left(t/\tau\right)} + Z_\text{min},
\end{equation}
where the $Z_\text{min}$ is the stellar metallicity at the initial onset of star formation and $Z_\text{max}$ is the stellar metallicity today. The share of weights in stellar metallicity at a given age is parametrized using the relation
\begin{equation}\label{eq:sim-kernel}
w_{ij} = (Z_i/Z_\text{peak})^2\exp{\left(-2Z_i/Z_\text{peak}\right)},
\end{equation}
where $Z_\text{peak}$ is the mode of the distribution and is given by the ChEH at any given time $t_j$ \citep[e.~g.,][]{Conroy2009}. To keep the analysis simple, we arbitrarily fix the initial onset of star formation to the age of the Universe $\sim14\,$Gyr and the corresponding initial stellar metallicity $Z_\text{min}=0.19\,Z_{\sun}$. This way, the only free parameters are the timescale $\tau$ and the $Z_\text{max}$, which would be related to the integrated stellar mass of galaxies through the stellar mass-metallicity relation \citep[e.~g.,][]{Gallazzi2005N, Panter2008}. Again, for the sake of simplicity we assume both these parameters ($\tau$ and $Z_\text{max}$) to be uncorrelated, so we sample the space of SFH and ChEHs randomly assuming uniform distributions in the ranges $[1, 20]\,$Gyr and $[0.25, 1.50]\,Z_{\sun}$, respectively. We are able to populate this parameter space by sampling $1000$ points. Finally to mimic the observables from this toy simulations, we use the same SSP models as in the spectral fitting method described in \S~\ref{sec:spectral-fitting} and added Gaussian random noise at different levels of signal-to-noise ratio, namely, $S/N=25,50,100,500$. In order to measure the stability of the results against the simulated noise, we draw for each SFH and ChEH $10$ noise realizations, so that we end up with $10\times1000$ spectra.

We are aware the suppositions made above are not the most realistic ones, but merely reflect some physically motivated expectancies, namely: \textit{(i)} the SFR should decrease in time due to local and global (self-)regulating processes and \textit{(ii)} the ChEH should be the cumulative history of heavy-elements production, assuming that gas inflows and outflows play no important r\^ole in this regard. We reckon these simulations are enough to demonstrate the relation between the second moment of the MDFs presented in this work, $\sigma_Z$, and the typical timescale of the SFH and ChEH, $\tau$.

We feed the simulated observations to the \textsc{FIT3D} algorithm to obtain the stellar decomposition $w_{ij}$ as described in \S~\ref{sec:spectral-fitting}. In Fig.~\ref{fig:sim-showcase}, top row, we show an example of these toy simulations, with $\tau=13.58\,$Gyr and $Z_\text{max}=1.16\,Z_{\sun}$. From left to right panel, we show the SFH, the ChEH and the $w_{ij}$ distributions at the different ages $t_j$. In the second row (from top to bottom), we show the simulated spectra (black) along with its $1\,\sigma$ (grey region), the best fitted average spectra (blue), shifted to increasing fluxes with increasing $S/N$ values for better visualization. In the third row, we compare directly the ChEH simulated (black) and its intrinsic spread (grey region) with the retrieved in (blue) crosses. Each cross is centred at the averaged ChEH, the vertical bar corresponds to the $1\,\sigma$ spread, both across noise realisations, the horizontal bar corresponds to the age bin width. To improve the visualization, each cross is shifted according to its $S/N$ value. In the bottom row, we show the resulting MDF for all $S/N$ values (blue) and the simulated one (black). It is clear that even though we successfully fit each instance of the spectrum for this particular simulation, we are not retrieving a perfect match of the SFH and ChEH. Yet, we are able to confidently retrieve the MDF when compared to the real one. Interestingly, the KDE-like method implemented seems to be also robust against the chosen shape of the kernel (Gaussian), given that we assumed a different shape (Eq.~\ref{eq:sim-kernel}) for the simulations. Notwithstanding, we also found that when $\tau$ is small ($\sim1\,$Gyr) the resulting MDF tends to be biased towards higher metallicities, even for the best $S/N$. It is also clear that for low values of $S/N=25,50$ (typical in the outskirts of disk-like galaxies) we can recover a bimodal MDF despite the unimodal behaviour of the real one.

\begin{figure}
\includegraphics[scale=0.45]{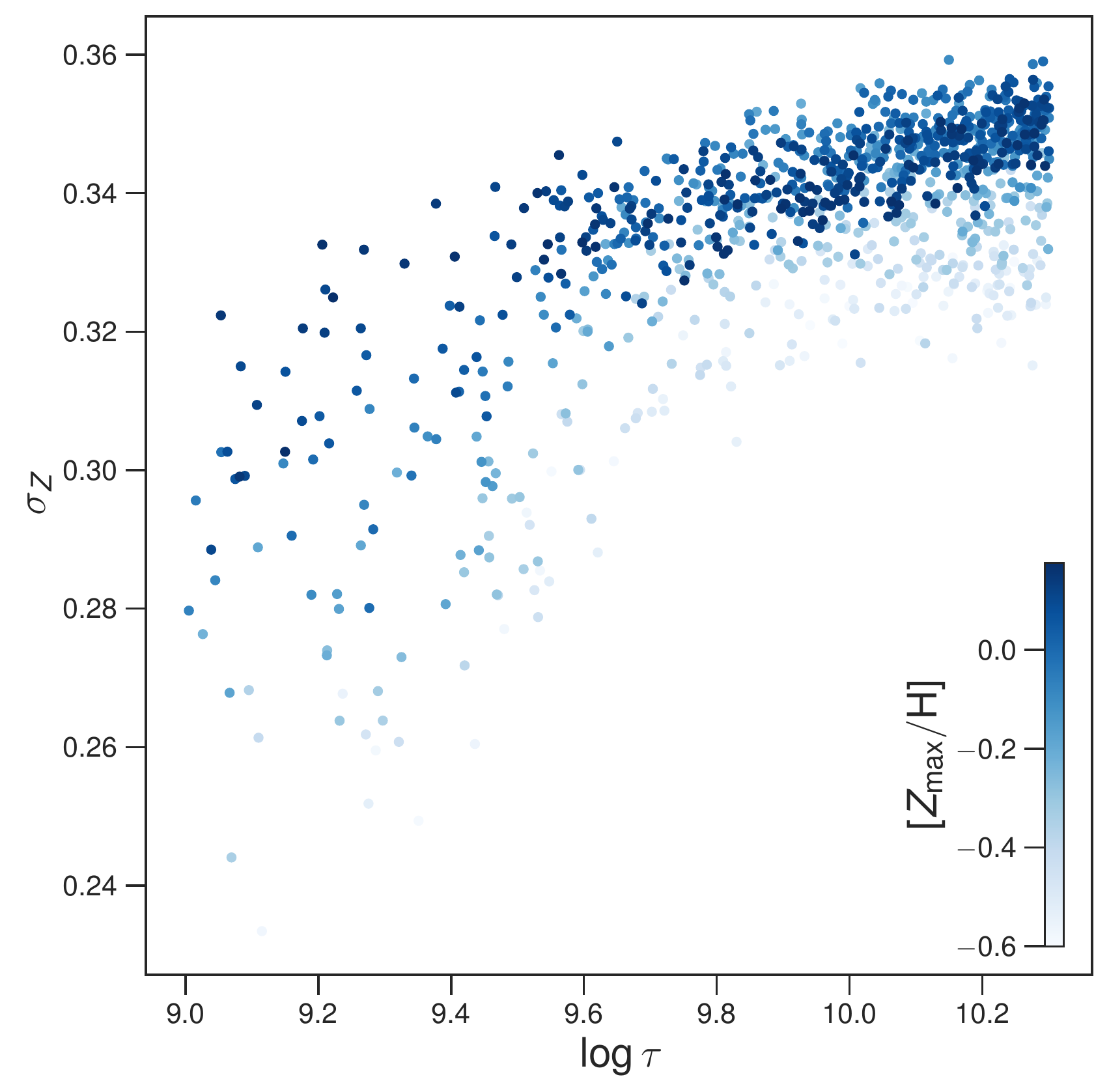}
\caption{The relation between the SFH and ChEH timescale, $\tau$ and the spread of the retrieved MDF as a function of the present-day stellar metallicity $Z_\text{max}$. The correlation is clear and in the same sense as the predicted using the real data set. It is clear also that the strength of such correlation may depend on other parameters and on the model adequacy.}
\label{fig:sim-tau-mdf}
\end{figure}

In Fig.~\ref{fig:sim-tau-mdf} we show the relation between $\tau$ and the $\sigma_Z$ of the retrieved MDF for the complete simulation set. In this plot, each point is computed from the averaged noise realizations for each SFH and ChEH. As predicted using the real data set, we can see a strong correlation between the two parameters, which scale is modulated by the present day stellar metallicity, $Z_\text{max}$. We conclude from these simulations that in general we can recover the MDF, although for rather simple yet physically motivated toy simulations. The fact that this happens regardless of the true shape of the kernel (c.~f., Eq.~\ref{eq:sim-kernel} \emph{versus} \ref{eq:mdf-kernel}) is encouraging. Further to this, we note that the range of $S/N$ at which these simulated MDFs are well recovered agrees with that of our original observed data. In the real data these $S/N$ values consist on the radial average (across $\sim100\,$spaxels) that individually contributed to only $S/N\sim20$. The final $S/N$ is in fact of the order of a few hundreds. In order fully to quantify the uncertainties in the MDF method introduced in this study, we will analyze more complex and realistic simulations in an upcoming work.

\update{\section{Stellar metallicity profiles}

\begin{figure*}
\includegraphics[scale=0.45]{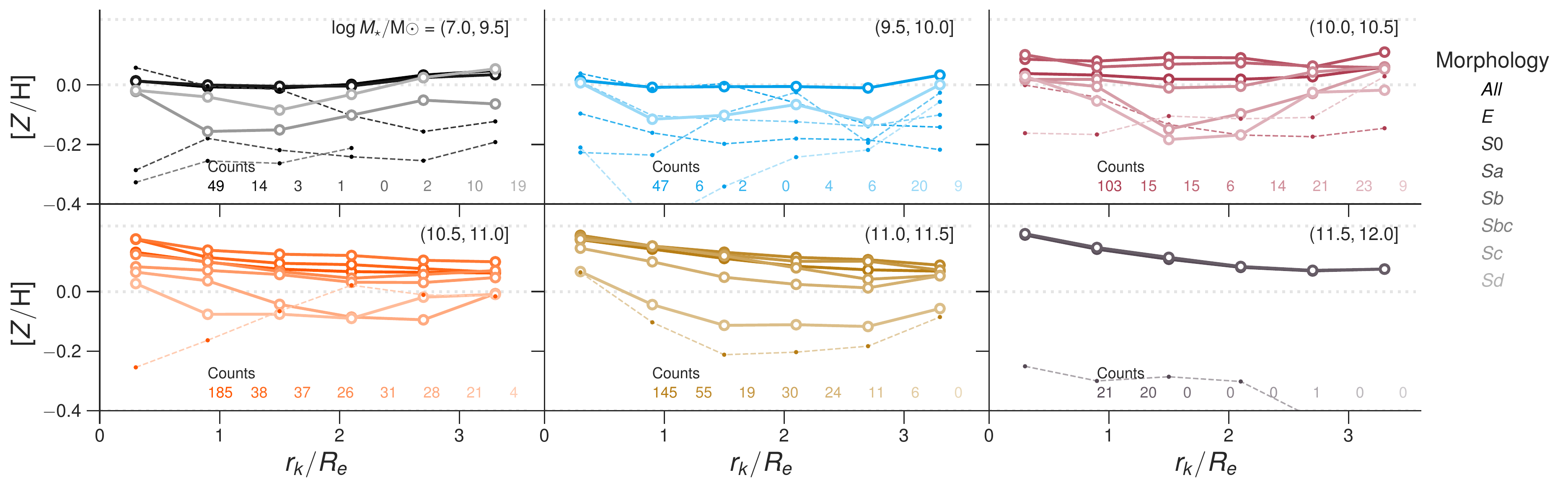}
\caption{\update{Similar to Fig.~\ref{fig:mdfs-profiles}, but segregating first by stellar mass (color coded) and then by morphological class. The profiles are drawn in fainter tones according to the morphological class, from earlier to later galaxy types.}}
\label{fig:mdfs-profiles-inverted}
\end{figure*}
In Fig.~\ref{fig:mdfs-profiles-inverted} we show the stellar metallicity profiles similar to Fig.~\ref{fig:mdfs-profiles}, but segregating first by stellar mass and then by morphological class. In each panel the profiles are drawn in progressively faded colour tone from earlier to later galaxy types. If we focus our attention on the well populated bins ($>10$ galaxies), the variance in the profiles increases among morphological classes with decreasing stellar mass, suggesting that morphology becomes more important in setting the stellar metallicity scale for low-mass galaxies. This r\^ole seems to be delegated to the stellar mass for massive galaxies.
}


\bsp	
\label{lastpage}
\end{document}